\documentclass{jfm}

\usepackage{graphicx}
\usepackage[percent]{overpic} 
\usepackage{dcolumn}
\usepackage{bm}
\usepackage{hyperref}

\usepackage{amsmath, amsfonts, amssymb, amscd} 
\usepackage{subfigure} 
\usepackage{color}
\usepackage[usenames,dvipsnames]{xcolor} 
\usepackage{xfrac}
\usepackage{float} 
\usepackage{cancel}
\usepackage{enumerate}
\usepackage[mathscr]{eucal}
\usepackage{setspace}
\usepackage{siunitx}
\usepackage{natbib}

\newcommand{\abs}[1]{\left| #1 \right|}
\newcommand{\Order}[1]{\mathcal{O}\left(#1\right)}

\renewcommand{\bar}[1]{\overline{#1}}

\newcommand{\intinf}{\int_{-\infty}^{\infty}}

\newcommand{\dop}[1]{\operatorname{d}\! #1}

\newcommand{\eps}{\varepsilon}
\newcommand{\lt}{\tilde{\lambda}}
\newcommand{\km}{k_-}

\newcommand{\kt}{\tilde{k}}
\newcommand{\ktp}{\tilde{k}_+}
\newcommand{\Ncal}{\mathcal{N}}
\newcommand{\om}{\omega}
\newcommand{\phib}{\overline{\phi}}
\newcommand{\pbmin}{\phib_{\mathrm{min}}}

\newcommand{\R}{\mathbb{R}}
\newcommand{\rmd}{\,\mathrm{d}}
\newcommand{\phit}{\tilde{\phi}}
\newcommand{\tht}{\tilde{\theta}}
\newcommand{\Tt}{\tilde{t}}
\newcommand{\ub}{\overline{u}}
\newcommand{\wt}{\tilde{\om}}
\newcommand{\zt}{\tilde{z}}
\newcommand{\At}{\tilde{a}}
\newcommand{\ifl}{^{(i)}}
\newcommand{\efl}{^{(e)}}

\newcommand{\Acal}{\mathcal{A}}

\DeclareSIUnit[number-unit-product = \,]{\pixel}{pixel}

\date{\today}

\title{Solitary wave fission of a large disturbance in a viscous fluid conduit}

\author{M. D. Maiden\aff{1},
        N. A. Franco\aff{1,2},
        E. G. Webb\aff{1},
        G. A. El\aff{3},
        \and \,
        M. A. Hoefer\aff{1}\corresp{\email{hoefer@colorado.edu}},
        }

\affiliation{\aff{1} Department of Applied Mathematics, University of Colorado Boulder,
Boulder, CO 80309, USA
\aff{2} Department of Physics, University of Alaska Fairbanks, Fairbanks, AK, 99775
\aff{3} Department of Mathematics, Physics and Electrical Engineering, Northumbria University, Newcastle upon Tyne, UK}

\begin{document}
\maketitle
\begin{abstract}
  This paper presents a theoretical and experimental study of the
  long-standing fluid mechanics problem involving the temporal
  resolution of a large, localised initial disturbance into a sequence
  of solitary waves. This problem is of fundamental importance in a
  range of applications including tsunami and internal ocean wave
  modelling. This study is performed in the context of the viscous
  fluid conduit system---the driven, cylindrical, free interface
  between two miscible Stokes fluids with high viscosity contrast.
  Due to buoyancy induced nonlinear self-steepening balanced by stress
  induced interfacial dispersion, the disturbance evolves into a
  slowly modulated wavetrain and further, into a sequence of solitary
  waves.  An extension of Whitham modulation theory, termed the
  solitary wave resolution method, is used to resolve the fission of
  an initial disturbance into solitary waves.  The developed theory
  predicts the relationship between the initial disturbance's profile,
  the number of emergent solitary waves, and their amplitude
  distribution, quantifying an extension of the well-known soliton
  resolution conjecture from integrable systems to non-integrable
  systems that often provide a more accurate modelling of physical
  systems.  The theoretical predictions for the fluid conduit system
  are confirmed both numerically and experimentally.  The number of
  observed solitary waves is consistently within 1--2 waves of the
  prediction, and the amplitude distribution shows remarkable
  agreement.  Universal properties of solitary wave fission in other
  fluid dynamics problems are identified.

  \end{abstract}


\maketitle

\section{Introduction}

A fundamental problem in fluid dynamics is the long-time resolution of
a large, localised disturbance.  In inviscid fluids, a prominent
feature of this resolution is the emergence of a solitary wavetrain.
This process is generally referred to as soliton fission and has been
observed in a variety of fluid contexts.  For example, while intense
earthquakes can lead to the vertical displacement of the ocean surface
by several meters, its horizontal extent can reach 10--100 kilometers
\cite{geist_implications_2007}, which, under appropriate shallowness
conditions, can evolve into a large number of surface solitary waves
\citep{matsuyama2007,arcas_seismically_2012}.  Another important
example is the generation of large amplitude internal ocean solitary
waves with two identified soliton fission mechanisms: 1) an initial,
broad displacement of internal temperature and salinity
\citep{osborne_internal_1980} and 2) the propagation of a large
internal solitary wave onto a shelf
\citep{farmer_generation_1999,vlasenko_tidal_2014}.  In both
scenarios, the result is the same---the generation of a large number
of rank-ordered solitary waves.  In fact, the well-known soliton
fission law by \cite{djordjevic_fission_1978} for scenario 2 was
obtained by modeling it with an initial, broad disturbance to the
constant coefficient Korteweg-de Vries (KdV) equation, a weakly
nonlinear, long wave model.  More generally, the disintegration of a
broad disturbance into solitary waves is the inevitable result of
boundary or topography interaction with an undular bore or dispersive
shock wave (DSW) that results from a sharp gradient due to a variety
of reasons \cite{el_dispersive_2016}.

Despite the prevalence of soliton fission in fluid dynamics, its
theoretical description has primarily been limited to completely
integrable partial differential equations (PDEs) such as the KdV
equation.  First attempts to understand this problem began with the
celebrated Zabusky-Kruskal numerical experiment of an initial cosine
profile for KdV \citep{zabusky_interaction_1965}.  Asymptotics of KdV
conservation laws
\citep{karpman_asymptotic_1967,johnson_development_1973} and the
inverse scattering transform
\citep{segur_korteweg-vries_1973,deng_small_2016} yield a prediction for the
number of solitons based on eigenvalue counting and an estimate for
the amplitudes of fissioned solitons from an initial profile.

Because of its ubiquity, we seek a deeper understanding of soliton
fission that results from a broad initial condition, hereafter
referred to as the box problem due to the initial profile's wide
shape.  A new method based on Whitham averaging theory
\citep{whitham_linear_1974} that does not require integrability was
first proposed and applied to the Serre/Su-Gardner/Green-Naghdi
equations for fully nonlinear shallow water waves in
\cite{el_asymptotic_2008} and, partially, to the defocusing nonlinear
Schr\"odinger equation with saturable nonlinearity in
\cite{el_theory_2007}. The method draws upon principles first
developed to describe DSWs that result from step initial data
\citep{el_resolution_2005}.  The long-time evolution into a solitary
wavetrain is one component of the soliton resolution conjecture, which
proposes that localised initial conditions to nonlinear dispersive
wave equations generically evolve into a soliton wavetrain and small
amplitude dispersive radiation, originally formulated within the
context of integrable PDEs such as the KdV equation
\citep{segur_korteweg-vries_1973,schuur_emergence_1986,deift_collisionless_1994}.
Because the method presented here is not reliant on integrability of
the underlying PDE and yields concrete predictions for the number of
solitary waves and their amplitude distribution that result from broad
initial disturbances, we refer to this approach as the solitary wave
resolution method.  An important feature of the solitary wave
resolution method is that it bypasses an analysis of the full Whitham
modulation equations---which are generally difficult to analyse---in
favor of the exact zero amplitude and zero wavenumber reductions of
the full Whitham equations that admit a general structure and form
that is amenable to further analysis.

We note that the solitary wave resolution method does not resolve the
second component of the soliton resolution conjecture---the small
amplitude dispersive radiation.  For sufficiently broad boxes, this
component of the conjecture is negligible, as is well-known for the
KdV equation (see, e.g.,
\cite{karpman_non-linear_1974,whitham_linear_1974}).  We quantify the
contributions of the radiation and solitary wave components
numerically for a specific initial disturbance in the conduit
equation.

Our basic hypothesis in this work is that a large initial disturbance
in certain nonintegrable equations (e.g., the conduit equation
described below) results most prominently in the fission of solitary
waves, which enables us to apply Whitham averaging theory
\citep{whitham_linear_1974}.  This hypothesis is motivated by rigorous
semiclassical analysis of the KdV equation \citep{lax_zero_1979-1} and
is confirmed by numerical simulations for the conduit equation.
Moreover, this hypothesis has been successfully applied to the
nonintegrable Serre equations \citep{el_asymptotic_2008}.

The solitary wave fission problem has been studied experimentally,
primarily in water wave tanks modeled by the KdV equation
\citep{hammack_korteweg_1974, hammack_korteweg_1978}.  More recent
water wave experiments physically recreated the Zabusky-Kruskal
numerical experiment, observing recurrence as well as soliton fission
\citep{trillo_experimental_2016}.  These experiments exhibit excellent
agreement with WKB theory applied to the inverse scattering transform
\citep{deng_small_2016}.  The WKB approach has also been applied to
the defocusing nonlinear Schr\"{o}dinger equation, yielding the number
and amplitudes of emergent solitons \citep{deng_recurrence_2017}.
However, none of these quantitative methods are applicable to
non-integrable equations.

\begin{figure}
  \centering
  \includegraphics{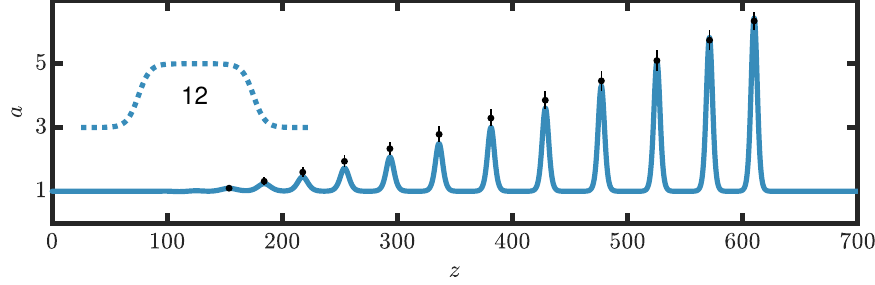}
  \caption{An example box initial condition (inset) and its long-time
    numerical evolution according to the conduit equation
    \eqref{eq:conduit}. The number in the inset denotes the predicted
    number of solitary waves from that initial condition based on the
    solitary wave resolution method, and the black circles with
    vertical bars denote the ranges from a quantiled distribution of
    the predicted solitary wave amplitudes, both derived later in this
    paper.}
  \label{fig:conduitIVP}
\end{figure}

This paper presents solitary wave fission experiments and modulation
theory for the interfacial dynamics between two high-viscosity,
miscible fluids, one rising buoyantly within another.  Original
experiments demonstrated that solitary waves preserve their shape and
form despite long distance propagation and interaction with other
solitary waves \citep{olson_solitary_1986,scott_observations_1986}.
In fact, in both of these experimental papers, solitary waves were
generated by the fission of a large, initial disturbance.  We apply
modulation theory for solitary wave fission introduced in
\citep{el_asymptotic_2008} to the box problem for a PDE model of this
fluid context known as the conduit equation
\cite{lowman_dispersive_2013}
\begin{equation}\label{eq:conduit}
  a_t + (a^2)_z - \left(a^2\left(a^{-1}a_t\right)_z\right)_z = 0 .
\end{equation}
In the derivation of the conduit equation, no restriction is placed on
the magnitude of the nondimensional, circular cross-sectional area
$a(z,t)$, where $z$, $t$ are the scaled height and time, respectively,
assumed to be much larger than the conduit diameter and a
characteristic advective time scale. This equation is also an
asymptotic, long-wave model of magma flows rising through the Earth's
mantle
\citep{barcilon_nonlinear_1986,whitehead_wave_1988,helfrich_solitary_1990}
and to a comparatively simple laboratory experiment
\citep{olson_solitary_1986,scott_observations_1986,whitehead_wave_1988,helfrich_solitary_1990,lowman_dispersive_2013,maiden_observation_2016,maiden_hydrodynamic_2018,anderson_controlling_2019}.
Equation \eqref{eq:conduit} fails the so-called Painlev\'e test for
integrability \citep{harris_painleve_2006} and has at least two
conservation laws \citep{harris_conservation_1996} therefore is an
excellent candidate to test the more broadly applicable solitary wave
resolution method for the initial value problem
consisting of \eqref{eq:conduit} and
\begin{equation}\label{eq:conduitIVP}
  \begin{array}{cc}
    a(z,0) = 1 + a_0(z), & \lim_{\abs{z}\to\infty} a_0(z) = 0 ,
  \end{array}
\end{equation}
where $a_0(z)$ is a broad, localised disturbance with exactly one
critical point at the maximum
\begin{equation}
  \label{eq:3}
  a_m = \max_{z \in \mathbb{R}} a_0(z) .
\end{equation}
Note that $a = a_m+1$ at the maximum, i.e., $a_m$ measures the
amplitude of the disturbance exceeding the unit background area ratio
$a = 1$.  We will quantify the profile's broadness more precisely
later on but for $a_0(z)$ in the shape of a box, then a sufficiently
wide box will do.  Formally, $a_0\in\mathcal{C}^\infty(\R)$ as well,
but the relaxation of this assumption still aligns with the
theoretical results.  We shall assume that the support of $a_0(z)$ is
$[-w,0]$ where $w>0$ is the box width.  The solitary wave resolution
method utilizes the characteristics of the Whitham modulation
equations to estimate the number of solitary waves and the solitary
wave amplitude distribution resulting from a large-scale initial
condition.  An example initial condition and its numerically evolved
state according to the conduit equation \eqref{eq:conduit} are shown
in figure \ref{fig:conduitIVP}.  The theoretically predicted solitary
wave number (12, derived in section~\ref{sec:numb-solit-waves}) is
correct and the predicted amplitudes fall well within the ranges
determined from the quantization of the continuous amplitude
distribution (derived in section~\ref{sec:distr-solit-wave}).

The conduit equation \eqref{eq:conduit} can be approximated by the KdV
equation
\begin{equation}
  \label{eq:KdV}
  u_\tau + uu_x + u_{xxx} = 0
\end{equation}
in the small-amplitude, long-wavelength regime with the scaling
\citep{whitehead_korteweg-devries_1986}
\begin{equation}\label{eq:KdV2conduit}
  \begin{array}{ccc}
    \tau = \delta^{3/2}t,
    & 
      x = \delta^{1/2}2^{-1/3}\left(z-2t\right),
    & 
      u = 2^{2/3}\delta(a-1), \quad 0 < \delta \ll 1,
  \end{array}
\end{equation}
where $\delta$ is a characteristic disturbance amplitude deviation
from unit background.  The formulae for the expected number of
solitons $\Ncal$ and the amplitude ($\Acal$) density function
$f(\Acal)$ for the initial profile $u(x,0) = u_0(x)$ to the KdV
equation \eqref{eq:KdV} are \citep{el_asymptotic_2008}
\begin{equation}
  \label{eq:NsoliKdV}
  \begin{split}
    \Ncal &= \frac{1}{\pi\sqrt{6}}\intinf \sqrt{u_0(x)}\rmd x, \\
    f(\Acal) &= \frac{1}{4\pi\sqrt{6}}\int_{x_1}^{x_2}\frac{\rmd
      x}{\sqrt{u_0(x)-\Acal/2}}, \quad 0 \le \Acal \le 2u_m .
  \end{split}
\end{equation}
Here, $x_1$ and $x_2$ are the intersections of the initial condition
$u_0$ with the value $\Acal/2$.  The initial data is assumed to be on
a zero background with maximum $u_m = \max u_0(x)$.  For initial data
consisting of a box of width $w$ and height $u_m$,
equation~\eqref{eq:NsoliKdV} becomes
\begin{equation}
  \label{eq:1}
  \begin{split}
    \mathrm{for} ~ u_0(x) &=
    \begin{cases}
      u_m & -w < x < 0 \\
      0 & \mathrm{else}
    \end{cases} , \quad \Ncal = \frac{w \sqrt{u_m}}{\pi\sqrt{6}}, \\
    \mathcal{F}(\Acal) &= \frac{1}{\Ncal} \int_0^{\Acal} f(\Acal')\,
    \mathrm{d} \Acal' = 1 - \sqrt{1 - \frac{\Acal}{2 u_m}}, \quad 0
    \le \Acal \le 2u_m ,
  \end{split}
\end{equation}
where $\mathcal{F}(\Acal)$ is the cumulative distribution function of
the soliton amplitudes normalised by the total number of solitons.
The number of solitons agrees with that obtained by IST-related
approaches \citep{karpman_asymptotic_1967,ablowitz_soliton_2009} in
its asymptotic regime of validity $\Ncal \sim w \sqrt{u_m} \gg 1$.
This modulation theory approach to solitary wave fission can be
applied to any dispersive nonlinear wave equation that admits a
Whitham modulation description
\citep{whitham_linear_1974,el_dispersive_2016}.  We identify certain
universal properties of solitary wave fission, including the
independence of the normalised cumulative distribution function on box
width (e.g., $\mathcal{F}(\Acal)$ is independent of $w$).  We also
predict the linear dependence on box width $w$ of the solitary wave
fission number $\Ncal$.
    
This paper continues with section \ref{sec:expt} where we present
viscous fluid conduit fission experiments.  Section \ref{sec:theory}
includes relevant background information on the conduit equation
(\ref{eq:conduit}).  In sections \ref{sec:numb-solit-waves} and
\ref{sec:distr-solit-wave}, we develop the solitary wave resolution
method to estimate the number of solitary waves and their amplitude
distribution.  In light of the developed modulation theory, we return
to the experiments in section \ref{sec:comp-exper}.  We wrap up with
concluding remarks in section \ref{sec:conclusion}.

\section{Observation of solitary wave fission}\label{sec:expt}

We motivate our analysis by first presenting viscous fluid conduit
experiments on solitary wave fission.

\subsection{Experimental setup}

\begin{figure}
  \centering \subfigure[][]{\includegraphics[trim=0 17 0
    10,clip]{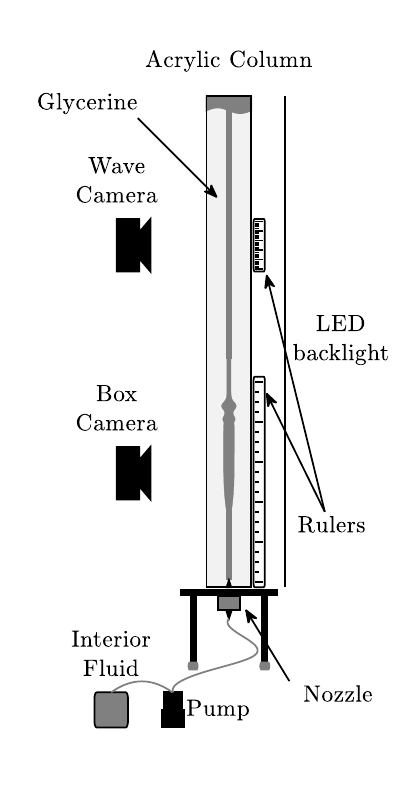}}
  \subfigure[][]{\includegraphics[trim=0 -40 25 0,clip]{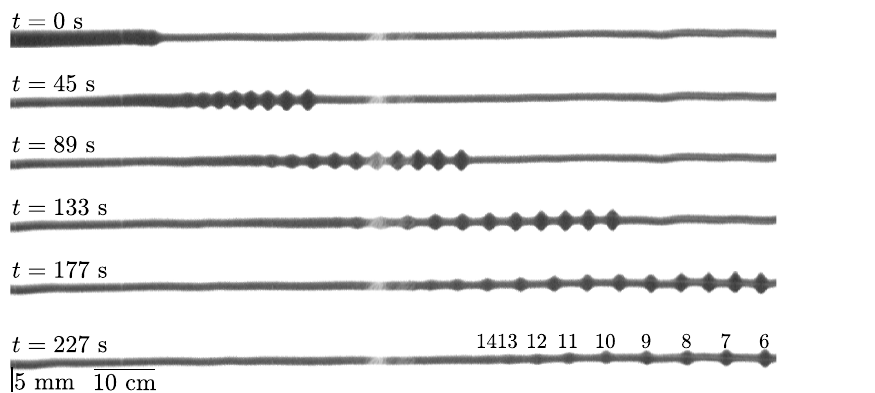}}
  \caption{(a) Schematic of the experimental apparatus. (b) Box (not
    entirely shown) with nominal width \SI{25}{\centi\meter} and total
    height (conduit diameter) \SI{3.2}{\milli\meter} at $t = 0$
    develops into a rank-ordered solitary wavetrain with 12--13
    visible solitary waves at $t = \SI{177}{\second}$.  The lead
    solitary wave (diameter $\SI{4.6}{\milli\meter}$) propagates on
    the background conduit with diameter $\SI{2.0}{\milli\meter}$.  At
    a later time ($t = \SI{227}{\second}$) the smallest 9 solitary
    waves---verified by zoomed-in images from the wave camera---are
    labelled by their amplitude ranking.  The $\ang{90}$ clockwise
    rotated images exhibit an 8:1 aspect ratio.  Slight discoloration
    near image centre is due to an external scratch. Measured
    experimental parameters:
    $\mu^{(i)} = \SI{4.95d-2}{\pascal\second}$,
    $\mu^{(e)} = \SI{1.0}{\pascal\second}$,
    $\rho^{(i)} = \SI{1.205}{\gram\per\cubic\centi\meter}$,
    $\rho^{(e)} = \SI{1.262}{\gram\per\cubic\centi\meter}$,
    $Q_0 = \SI{0.25}{\cubic\centi\meter\per\minute}$.  The Poiseulle
    flow relations \eqref{eq:19} and \eqref{eq:20} yield
    $2R_0 = \SI{2.0}{\milli\meter}$ and
    $U_0 = \SI{1.35}{\milli\meter\per\second}$.}
  \label{fig:expt_setup}
\end{figure}

The experimental setup is nearly identical to that used by
\cite{anderson_controlling_2019} and consists of a square acrylic
column with dimensions $\SI{4}{\centi\meter} \times
\SI{4}{\centi\meter} \times \SI{200}{\centi\meter}$, filled with
glycerine, as shown in figure \ref{fig:expt_setup}(a).  The interior
fluid (identified by the superscript $^{(i)}$) consists of certain
ratio of glycerine, water, and black food coloring, which is injected
through a nozzle installed at the column's base.  The ratio is chosen
so that the interior fluid has both lower density, $\rho^{(i)} <
\rho^{(e)}$, and significantly lower viscosity, $\mu^{(i)} \ll
\mu^{(e)}$, than the exterior fluid denoted by the superscript
$^{(e)}$.  Miscibility of the two fluids implies that surface tension
effects are negligible.  The nominal parameter values used in the
experiments presented here are those in table \ref{tab:physQuants}.
\begin{table}
  \centering
  \begin{tabular}{cl}
    $\mu^{(i)}$        & \SI{3.66d-2}{\pascal\second} \\
    $\mu^{(e)}$        & \SI{1.296}{\pascal\second}   \\ 
    $\rho^{(i)}$       & \SI{1.198}{\gram\per\cubic\centi\meter} \\ 
    $\rho^{(e)}$       & \SI{1.260}{\gram\per\cubic\centi\meter}  \\[0.5mm]
    $\eps$            & 0.0283             \\ 
    $Q_0$             & \SI{0.50}{\cubic\centi\meter\per\minute}\\[0.5mm]
    $2R_0$            & \SI{2.1}{\milli\meter}\\
    $U_0$            & \SI{2.3}{\milli\meter\per\second}\\
  \end{tabular}
  \caption{Densities, viscosities, viscosity ratio, background flow
    rate $Q_0$, and associated background conduit diameter $2R_0$,
    mean flow rate $U_0$ according to equations~\eqref{eq:19},
    \eqref{eq:20}, respectively for the reported experiments (except
    figure~\ref{fig:expt_setup}(b)).}
  \label{tab:physQuants}
\end{table}
    
A high precision, computer-controlled piston pump is used to inject
the interior fluid with a pre-determined temporal flow profile.
Buoyancy and steady injection at a fixed volumetric flow rate ($Q_0$
in table \ref{tab:physQuants}) leads to a vertically uniform fluid
conduit, which is referred to as the background conduit, and is
verified to be well-approximated by the pipe (Poiseuille) flow
relation (see, e.g., the supplementary material in
\cite{maiden_observation_2016})
\begin{equation}
  \label{eq:19}
  2 R_0 = \left ( \frac{2^7 \mu^{(i)} Q_0}{\pi g (\rho^{(e)} -
      \rho^{(i)})} \right )^{1/4}, 
\end{equation}
where $2 R_0$ is the conduit diameter and $g$ is the acceleration due
to gravity.  The mean vertical advective velocity within the conduit
according to pipe flow is
\begin{equation}
  \label{eq:20}
  U_0 = \frac{g R_0^2(\rho^{(e)} - \rho^{(i)})}{8\mu^{(i)}} .
\end{equation}

Data acquisition is performed using high resolution digital cameras
equipped with macro lenses, one to capture the initial box profile and
one near the top of the apparatus (the far-field) to capture the
solitary wavetrain.  A ruler is positioned beside the column within
camera view for calibration purposes in order to quantify the observed
box width.  All amplitudes (solitary wave and box) are reported as
cross-sectional areas that are normalised to the observed mean
background cross-sectional area.  This facilitates future comparison
with theoretical results for the nondimensional conduit equation
\eqref{eq:conduit}.

\subsection{Methods}\label{sec:Methods}

We use the characteristic control method described in
\citep{anderson_controlling_2019} to generate a volumetric flow rate
profile that results in a box-like structure in the lower part of the
column with a pre-specified width $w$ and nondimensional
cross-sectional area $a_m+1$.  Figure \ref{fig:expt_setup}(a) displays
a schematic of the experiment and figure~\ref{fig:expt_setup}(b)
depicts the experimental time development of a box-like profile.  The
lower ``box camera'' takes several images before, during, and after
the predicted box development time.  After the leading edge of the box
forms, the pump rate is quickly reduced to the background rate $Q_0$,
and the box evolves into oscillations that rise up the conduit.  Once
the leading oscillation reaches the upper ``wave camera'' imaging
window, images are taken at \SI{0.2}{\hertz} for several minutes, to
ensure that all waves originating from the box have had sufficient
time to propagate through the viewing window.  For the experiment
reported only in figure~\ref{fig:expt_setup}(b), an additional camera
(not shown in figure~\ref{fig:expt_setup}(a)) is employed to image a
$\SI{1.25}{\meter}$ section of the column.  The large aspect ratio of
the full columnar dynamics imply the relatively low image resolution
of $\SI{43}{\pixel\per\centi\meter}$.  These dynamics will be directly
compared with the evolution predicted by the conduit equation in
Section \ref{sec:comp-exper}.
        	
The approximately white background and the opaque, black
conduit yield sufficient contrast for edge detection by identifying
the two midpoints between the maximum and minimum of the spline
interpolated horizontal image intensity.  The edge data is then
processed with a low-pass filter to reduce pixelation noise and the
effects of impurities in the exterior fluid.  The number of pixels
between the two edges is identified as the conduit diameter, which is
squared and normalised by the squared observed background conduit
diameter to obtain the dimensionless cross-sectional area $a$.  Our
imaging setup at both the ``box camera'' and ``wave camera'' in
figure~\ref{fig:expt_setup}(a) admit resolutions of 300 pixel
cm$^{-1}$ and between 132--228 pixel cm$^{-1}$, respectively
(generally higher resolution for smaller
boxes).  

We use the lower camera to determine the box shape. Note that near the
point of breaking, dispersion is no longer negligible; as a result, a
pure box is difficult to realize in the conduit system.  We use the
characteristic control method presented in
\citep{anderson_controlling_2019} to extract the time of box profile
formation, and use the nondimensionalized version of that profile as
the initial condition in further analyses of the conduit equation. An
example experimental box profile is shown in the $t=0$ panel of
figure~\ref{fig:expt_setup}(b).

For the upper camera, a wave-tracking algorithm is utilised to follow
all wave peaks across the imaging window.  Each candidate peak's
amplitude and position are validated against the conduit equation's
solitary wave speed-amplitude relation \cite{olson_solitary_1986}
\begin{equation}
  \label{eq:2}
  c(a_s) = \frac{2a_s^2\log{a_s} -a_s^2 + 1}{(a_s-1)^2} , 
\end{equation}
during the temporal window that the peak is in view.  The elevation
solitary wave amplitude $a_s > 1$ is measured from zero area, hence
must be larger than the background area $a = 1$.  Since solitary waves
exhibit the speed lower bound $c(a_s) > 2$, any observed wave peak
with a slower speed was discarded as small amplitude, dispersive wave
phenomena.

\subsection{Results}\label{sec:results}

\begin{figure}
  \centering
  \subfigure[]{\includegraphics{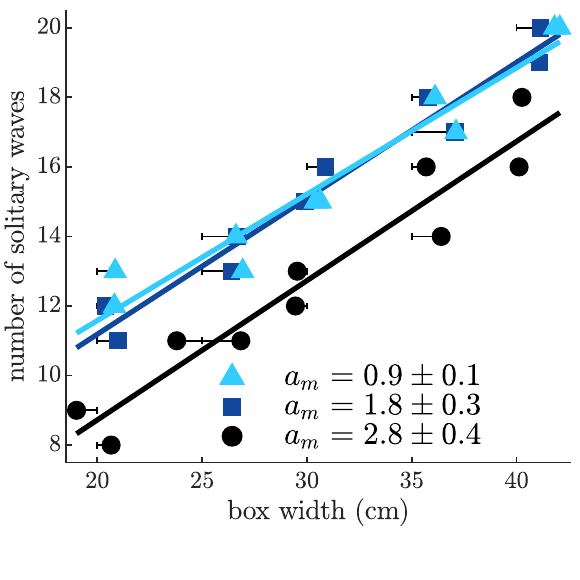}}
  \subfigure[]{\includegraphics{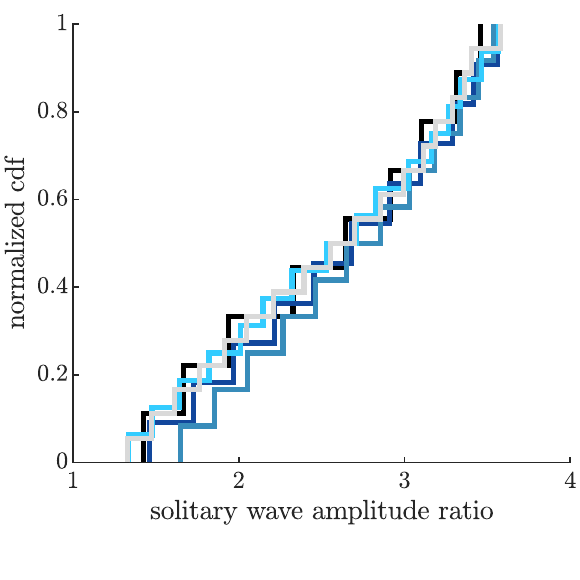}}
  \subfigure[]{\includegraphics{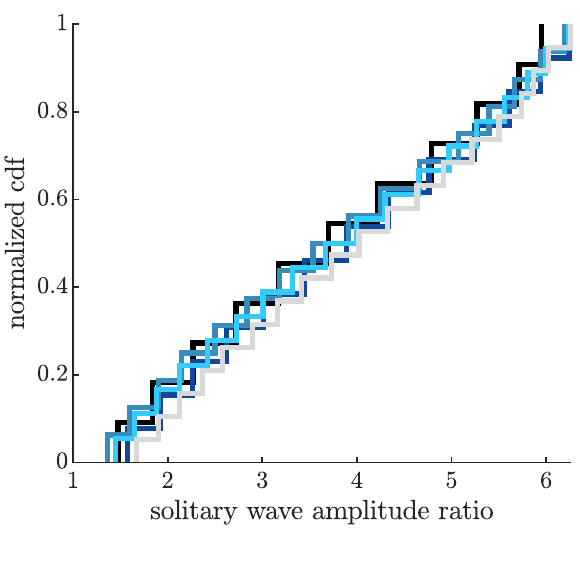}}
  \subfigure[]{\includegraphics{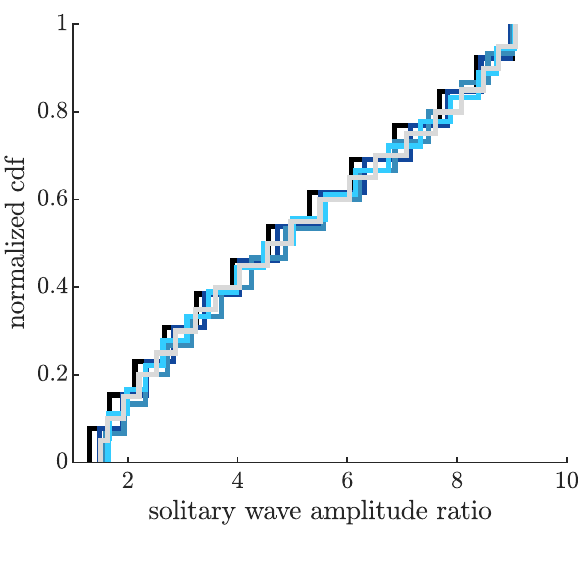}}
  \caption{(a) Observed number of solitary waves $\Ncal$ as a function
    of nominal dimensional box width and nondimensional box height
    area ratio $a_m$.  Linear fits for fixed box height are included.
    (b,c,d) Observed, normalised solitary wave cumulative distribution
    functions $\tilde{F}(\Acal)$, as defined in equation~\eqref{eq:4}.
    Each of (b), (c), and (d) correspond to nominal box heights
    $a_m = 1$, $a_m = 2$, and $a_m = 3$, respectively. In each panel,
    the normalised cdfs for the box widths $\{20,25,30,35,40\}$ cm are
    shown, with lighter linestyles corresponding to wider boxes.}
  \label{fig:solitonNum}
\end{figure}

A total of 30 experimental trials were executed according to the
protocol described in this Section with the experimental parameter
values identified in table~\ref{tab:physQuants}.  We generated 15 box
geometries and carried out 2 trials per geometry with the nominal
nondimensional box heights $a_m \in \{1,2,3\}$ and nominal box widths
$\{20,25,30,35,40\}$ cm---corresponding to nondimensional widths $w
\in \{90,112,134,156,178\}$ in the conduit equation \eqref{eq:conduit}
(the nondimensionalisation will be provided in Section
\ref{sec:cond-equat-backgr}).  The results are shown in figures
\ref{fig:solitonNum}(a-d).  Figure \ref{fig:solitonNum}(a) depicts the
observed number of solitary waves as functions of box width and
nondimensional height $a_m$.  Across all trials, between 8 and 20
solitary waves were generated, placing these fission experiments in
the large number of solitary waves regime, as expected for broad
initial conditions.  For fixed box height, the data exhibit an
approximately linear increase with width, as shown by the linear fits
for fixed $a_m$ in figure \ref{fig:solitonNum}(a).  Figures
\ref{fig:solitonNum}(b-d) report the normalised cumulative
distribution functions (cdfs) of solitary wave amplitudes.  Each panel
(b-d) includes the normalised cdfs for trials with a common box height
value $a_m$.  The normalised cdf $\mathcal{F}(\Acal)$ depends
parametrically on box width $w$ and box height $a_m$ and is defined as
\begin{equation}
  \label{eq:4}
  \mathcal{F}(\Acal) = \frac{\text{number of solitary
      waves with amplitudes}~a_s~\text{satisfying}~a_s \le
    1+\Acal}{\text{total number of solitary waves}} .
\end{equation}
While the staircase cdfs plotted in figures \ref{fig:solitonNum}(b-d)
have different shapes for different box heights, the box width
dependence of the normalised cdfs for fixed box height---i.e., for
each fixed panel---shows little variation.

The rest of this paper is concerned with developing a modulation
theory description for solitary wave fission in the conduit equation
box problem (equations~\eqref{eq:conduit} and \eqref{eq:conduitIVP}).  Our
analysis results in explicit predictions for the number of solitary
waves---linearly dependent on box width---and the normalised amplitude
cdf---independent of box width.  Following our analysis, we will
reconsider the experiments presented here.

\section{Conduit equation background}
\label{sec:cond-equat-backgr}

\label{sec:theory}


The conduit equation \eqref{eq:conduit} describes the dynamics of the
free interface between two viscous fluids: a highly dense, highly
viscous exterior fluid, and a less dense, less viscous interior fluid.
As the interior fluid is pumped steadily through the exterior fluid,
the interface resembles a deformable pipe whose walls are the
two-fluid boundary.  The circular cross-sectional area $A$ of this
pipe can be modeled as a function of time $T$ and vertical distance
$Z$ by the dimensional conduit equation
\citep{olson_solitary_1986,lowman_dispersive_2013}
\begin{equation}
  \label{eq:conduitDIM}
  A_T + \frac{g\Delta}{8\pi\mu\ifl}\left(A^2\right)_Z -
  \frac{\mu\efl}{8\pi\mu\ifl}\left(A^2\left(A^{-1}A_T\right)_Z\right)_Z
  = 0, 
\end{equation}
when $\eps=\frac{\mu\ifl}{\mu\efl}$, the interior to exterior dynamic
viscosity ratio, is small, $\Delta=\rho\efl-\rho\ifl$ is the
difference in exterior to interior fluid densities, and $g$ is
gravitational acceleration.  Equation \eqref{eq:conduitDIM} results
from the interplay between interior fluid buoyancy and continuity of
both the fluid velocity and interfacial stress at the two fluid
boundary.  This nonlinear dispersive partial differential equation is
a long-wave, slowly varying asymptotic reduction of the Navier-Stokes
equations for two fluids.  Restrictions include sufficiently small
Reynolds number and small interfacial steepness but there is no
restriction on the conduit amplitude, hence the dispersive term is
nonlinear \citep{lowman_dispersive_2013}.  The nondimensional form of
\eqref{eq:conduitDIM} is \eqref{eq:conduit}, obtained via the scalings
(c.f.~\eqref{eq:19}, \eqref{eq:20})
\begin{equation}\label{eq:scaling1}
    a=\frac{1}{\pi R_0^2}A, \quad z=\frac{\sqrt{8\eps}}{R_0}Z,
    \quad t = \frac{\sqrt{8\eps} U_0}{R_0}T.
\end{equation}
This transformation rescales the background conduit area of radius
$R_0$ to unity.  The conduit equation has been shown to admit a
variety of multiscale coherent wave solutions
\citep{maiden_modulations_2016}.
        
Previous experimental comparisons to dynamics predicted by the conduit
equation include solitary waves \citep{olson_solitary_1986}, their
interactions with each other
\citep{helfrich_solitary_1990,lowman_interactions_2014}, and
interactions with a dynamically changing mean flow
\citep{maiden_observation_2016,maiden_hydrodynamic_2018}.  Solitary
wave solutions can be obtained from the ordinary differential equation
that results from the travelling wave ansatz $a(z,t) = f(z-ct)$, where
the solitary wave speed $c$ is related to its total amplitude $a_s$
(measured from $a = 0$) on the background $\phib$ by the
speed-amplitude relation \citep{olson_solitary_1986}
\begin{equation}\label{eq:conduitsoliampl}
  c_s(a_s,\phib) =
  \frac{\phib\left(2a_s^2\left(\log{a_s}-\log{\phib}\right)
      -a_s^2 + \phib^2 \right)} {(a_s-\phib)^2}.
\end{equation}
    
Dispersive shock waves have also been studied theoretically and
experimentally in the viscous fluid conduit system
\citep{lowman_dispersive_2013-1,maiden_observation_2016,maiden_hydrodynamic_2018}.
Dispersive shock waves are the result of a sustained, large increase
in background conduit area from 1 to $\phib_- > 1$ and can be
characterised by a modulated periodic travelling wave solution of the
conduit equation, i.e., a solution of the form
\begin{equation}\label{eq:ModPeriodicWave}
  \begin{array}{ccc}
    a(z,t) = \phi(\theta), & \theta=kz-\om t,
    & \phi(\theta+2\pi) =  \phi(\theta).
  \end{array}
\end{equation}
Inserting this ansatz into equation \eqref{eq:conduit} and integrating
twice results in \citep{olson_solitary_1986}
\begin{equation}\label{eq:conduitPeriodic}
  (\phi ')^2 = g(\phi) = -\frac{2}{k^2}\phi -\frac{2}{\om
    k}\phi^2\log\phi + C_0 + C_1\phi^2,
\end{equation}
where $C_0$ and $C_1$ are real constants of integration.  The right
side of the equation can have up to three roots,
$\phi_1\leq\phi_2\leq\phi_3$, which parameterise the solution.
        
A physically relevant parameterisation of the periodic wave
$\phi(\theta)$ is given by three constants: the wavenumber $k$, the
wave amplitude $\Acal$ (defined as the difference between the wave's
maximum and minimum), and the wave mean $\phib$, which can be written
in terms of $C_0$, $C_1$, and $k$, or equivalently, in terms of
$\phi_j$, $j=1,2,3$.  The wave frequency is determined by the $2\pi$
periodicity of $\phi(\theta)$ as $\omega=\omega(k,\phib,\Acal)$.  The
modulation theory description of a DSW is achieved by allowing the
periodic wave's parameters to vary slowly relative to the wavelength
$2\pi/k$ and period $2\pi/\om$ while introducing the generalised
wavenumber $k = \theta_x$ and frequency $\om = -\theta_t$
\citep{lowman_dispersive_2013-1}. Then a DSW can be viewed as
connecting two distinguished limits of these modulated wave
parameters: the zero amplitude limit as $\Acal\to 0$ and the zero
wavenumber limit $k\to 0$.  When $\Acal\to 0$, the DSW solution limits
to small amplitude harmonic waves with the linear dispersion relation
\begin{equation}\label{eq:lineardispreln}
  \omega_0(k,\phib) = \frac{2k\phib}{1+k^2\phib}.
\end{equation}
When $k\to 0$, the DSW solution limits to a solitary wave that
satisfies the speed-amplitude relation \eqref{eq:conduitsoliampl}.
        
Allowing for slow modulations of $\phib$, $k$, and $\Acal$ in space
and time results in the conduit-Whitham equations.  The
conduit-Whitham equations consist of the conservation of waves $k_t +
\om_x = 0$, resulting from $\theta_{tx} = \theta_{xt}$, and the
averaging of the conduit equation's two conservation laws
\citep{barcilon_nonlinear_1986}
\begin{equation}\label{eq:ConsLaws}
  \begin{cases}
    a_t + (a^2-a^2(a^{-1}a_t)_z)_z= 0, \\[0.25em]
    \displaystyle \left(\frac{1}{a} + \frac{a_z^2}{a^2}\right)_t +
    \left(\frac{a_{tz}}{a}-\frac{a_z a_t}{a^2}-2\ln{a}\right)_z =0
  \end{cases}
\end{equation}
over the periodic wave family.  Using the following notation for
averaging over a wave period
\begin{equation}\label{eq:averaging}
  \bar{F} = \frac{1}{2\pi}\int_0^{2\pi} F(\theta)\dop{\theta},
\end{equation}
the Whitham equations are \citep{maiden_modulations_2016}
\begin{equation}\label{eq:whitham}
  \begin{cases}
    \phib_t + \left(\bar{\phi^2}-2k\omega\bar{\phi_\theta^2}\right)_z = 0\\
    \left(\bar{\frac{1}{\phi}}+k^2\bar{\frac{\phi_\theta^2}{\phi^2}}\right)_t
    -  2 \left(\bar{\ln{\phi}}\right)_z = 0\\
    k_t + \omega_z = 0
  \end{cases},
\end{equation}
where $\omega=\omega(k,\phib,\Acal)$ is the nonlinear wave frequency.
That the averaging operator \eqref{eq:averaging} approximately
commutes with partial differentiation is a result of scale separation
between the modulation---which is large and slow---and the periodic
wave's much shorter and faster spatial wavelength and temporal period,
respectively \citep{whitham_non-linear_1965,whitham_linear_1974}.

We remark that a rigorous, necessary condition for the stability of
conduit periodic waves is the hyperbolicity of the conduit-Whitham
equations \citep{johnson_modulational_2019}.  The conduit-Whitham
equations are known to be hyperbolic in an amplitude/wavenumber
dependent regime of phase space \citep{maiden_modulations_2016} and we
will operate within this regime.
        
If a certain self-similar, simple wave solution to the conduit-Whitham
equations exists (a 2-wave \citep{el_dispersive_2016}), we can obtain
expressions for the leading (solitary wave) and trailing (harmonic)
edge speeds in terms of the DSW jump parameter $\phib_-$, labeled
$s_+$ and $s_-$, respectively \citep{lowman_dispersive_2013-1}
\begin{equation}\label{eq:dswSpeeds}
  s_+ = \sqrt{1 + 8 \phib_-}-1, \quad s_- = 3 + 3 \phib_- -
  3\sqrt{\phib_-(8 + \phib_-)}.
\end{equation}
The solitary wave amplitude $a_+$ is implicitly determined by equating
$s_+$ with the solitary wave speed-amplitude relation
\eqref{eq:conduitsoliampl}
\begin{equation}
  \label{eq:22}
  c_s(a_+,1) = \sqrt{1 + 8 \overline{\phi}_-} - 1 .
\end{equation}
The trailing edge small amplitude wavepacket is characterized by the
wavenumber $k_-$, explicitly determined by equating the linear group
velocity $\partial_k \om_0$ to $s_-$
\begin{equation}\label{eq:DSWTrailingWavenum}
  k_-^2 = \frac{1}{4}\left(1 - \frac{4}{\phib_-} +
    \sqrt{\frac{1}{\phib_-}(8+\phib_-)}\right).
\end{equation}
        
The group velocity of the harmonic edge is always less than the speed
of the solitary wave edge.  Thus, a DSW in the conduit system is
always led by a solitary wave, with a trailing, continually expanding,
oscillating wavetrain that can exhibit backflow and instabilities for
sufficiently large jumps $\phib_-$
\citep{lowman_dispersive_2013-1,maiden_modulations_2016}.

We now return to the initial value problem \eqref{eq:conduitIVP} for
the conduit equation \eqref{eq:conduit} and our development of the
solitary wave resolution method.  Initially, the edges of the wide box
\eqref{eq:conduitIVP} can be treated as two well-separated, step-like
(Riemann) initial value problems.  As such, the rightmost edge will
evolve similar to a DSW, and the leftmost edge similar to a
rarefaction wave (RW).  However, finite box extent necessarily
implies the eventual interaction of the DSW and RW
\citep{el_generation_2002,ablowitz_soliton_2009}.  Ultimately, a finite solitary
wavetrain emerges from this interaction process.  We now use a
modification of conduit DSW theory \citep{lowman_dispersive_2013-1} to
determine the properties of this solitary wavetrain by applying the
solitary wave resolution method originally developed in
\citep{el_asymptotic_2008}.

\section{Number of solitary waves}
\label{sec:numb-solit-waves}

In what follows, we make the assumption that the box initial value
problem \eqref{eq:conduitIVP} for equation \eqref{eq:conduit} will
result in a slowly modulated wavetrain that can be described by the
Whitham modulation equations \eqref{eq:whitham}.  This assumption is
the cornerstone of the solitary wave resolution method
\citep{el_asymptotic_2008} and will later be verified by numerical
simulations.

Allowed to evolve long enough, the individual wave crests resulting
from the box initial conditions will separate with minimal overlap,
i.e., will result in a non-interacting solitary wavetrain.  To count
these waves, note that they are separated by exactly their wavelength,
defined in terms of the wavenumber as $2\pi/k$.  Consequently,
$k/2\pi$ is a wave crest density and we determine the total number of
waves $\Ncal$ in a wavetrain at time $t$ by
\begin{equation}\label{eq:Nsoli}
  \Ncal = \frac{1}{2\pi}\int_{-\infty}^\infty k(z,t)\rmd z. 
\end{equation}
This integral is finite at $t=0$ because the initial disturbance
$a_0(z)$ has compact support, implying $k\to 0$ as $\abs{z}\to\infty$
sufficiently fast.  The conservation of waves equation in the
conduit-Whitham modulation equations \eqref{eq:whitham} implies that
$\Ncal$ is independent of time.  Then the total number of fissioned
solitary waves that emerge in long time can be determined by the
wavenumber function $k(z,0)$ associated with the initial condition.
The challenge is to determine $k(z,0)$ when the waves are initially so
densely packed that there are no visible oscillations, i.e., the wave
amplitude $\Acal = 0$ and there is only the nonzero mean $\phib(z,0)$.
        
Whitham modulation theory can now be utilised to find a relationship
between the initial condition---the non-oscillatory data
\eqref{eq:conduitIVP} equates to the initial mean
$\phib(z,0) = 1+a_0(z)$ in modulation theory---and the wavenumber $k$.
For this, we note that the conduit-Whitham equations
\eqref{eq:whitham} are supplemented by conditions that ensure
continuity of the modulation solution at the trailing and leading
edges of the oscillatory wavetrain for all $t$. There exist only two
ways for the modulation solution to continuously match to the solution
of the dispersionless conduit equation
\begin{equation}\label{eq:hopf}
  \beta_t + 2\beta\beta_z = 0 .
\end{equation}
Either $k\to 0$ or $\Acal \to 0$.  The case $k \to 0$ is the solitary
wave limit and $\Acal \to 0$ is the small amplitude, harmonic wave
limit.  These limits are important for the modulation solution of a
DSW, with $\Acal \to 0$ at the leftmost, trailing edge and $k \to 0$
at the rightmost, leading edge.  Because early to intermediate time
evolution leads to the generation of a DSW, we identify these edges as
$z_-(t)$ and $z_+(t)$, respectively, and require
\begin{equation}\label{eq:wavetrainEdges}
  \begin{array}{lll}
    z = z_-(t): & \Acal=0, & \phib = \beta_-(t), \\
    z = z_+(t): & k=0,     & \phib = \beta_+(t)=1 ,
  \end{array}
\end{equation}
where the wave mean $\phib$ matches to the solution $\beta(z,t)$ of
the dispersionless conduit equation \eqref{eq:hopf} subject to the
initial condition $\beta(z,0) = 1+a_0(z)$ (cf.~\eqref{eq:conduitIVP}).
Then, $\beta_\pm(t) = \beta(z_\pm(t),t)$.  Consequently, equation
\eqref{eq:hopf} is valid outside the oscillatory region influenced by
the disturbance, i.e., for
$z \in (-\infty,z_-(t))\cup(z_+(t),\infty)$.  We note that the
dispersionless conduit equation \eqref{eq:hopf} (buoyancy driven flow
with negligible curvature induced interfacial stress) has been
experimentally shown to be a good approximation to the physical
conduit system when there are no oscillations, i.e., when the
interface is slowly varying \citep{anderson_controlling_2019}.  The
edges $z_{\pm}(t)$ in the boundary matching problem
\eqref{eq:whitham}, \eqref{eq:wavetrainEdges} with
$\beta_- \ne \hbox{const}$ can be determined by a recent extension of
the DSW fitting method developed by
\cite{kamchatnov_dispersive_2019}. This determination won't be
necessary in our construction in which we seek the long-time solitary
wave resolution.
        
When $\Acal\to0$, the vanishing oscillations do not contribute to the
averaging \eqref{eq:averaging}, so $\bar{F(\phi)}=F(\phib)$, for any
differential or algebraic operator $F$ \citep{el_dispersive_2016}.
Thus all $\theta$ derivatives of $\phi$ average to zero.  In this
case, the first and second conduit-Whitham equations
\eqref{eq:whitham} limit to the dispersionless conduit equation
\eqref{eq:hopf} but the conservation of waves modulation equation
remains and the wave frequency is the linear dispersion relation
\eqref{eq:lineardispreln} so that the modulation system reduces to
\begin{align}
  &\Acal=0: \nonumber \\
  \label{eq:reducedwhitham}
          & \phib_t + 2\phib\, \phib_z = 0, \\
  \label{eq:5}
          & k_{t} + \left(\omega_0(k,\phib)\right)_z = 0. 
\end{align}
Since the disturbance is initially non-oscillatory, we have
$\phib(z,0)=1+a_0(z)$, $z\in\mathbb{R}$ (c.f. \ref{eq:conduitIVP}).
However, because there are no initial oscillations, the initial
wavenumber is not well-defined.  We must appeal to properties of the
disturbance's evolution in order to uniquely define $k(z,0)$.  We do
so by identifying a simple wave relationship $k=k_-(\phib)$ between
the wavenumber and mean so that $k(z,0)=k_-(\phib(z,0))$.  The
rationale for the use of the simple wave relation is detailed in
\citep{el_asymptotic_2008} and is based on the fact that the DSW
trailing edge is a characteristic.
Equations~\eqref{eq:reducedwhitham} and \eqref{eq:5} have two
characteristic families
\begin{equation}
  \frac{dz}{dt} = 2\phib \quad \mathrm{and} \quad \frac{dz}{dt} = \omega_{0,k} .
\end{equation}
The first family corresponds to the decoupled evolution of the mean
flow equation \eqref{eq:reducedwhitham} and coincides with the slowly
varying evolution of the disturbance, e.g., the initial RW.  The
second family coincides with the vanishingly small amplitude
oscillations emerging from the edge of the evolving disturbance with
an envelope that moves with the group velocity.  It is the second
characteristic family that captures the evolution of the emergent
solitary wavetrain.  In order to obtain the relationship between $k$
and $\phib$ along the second characteristic family, we make the
simple wave ansatz $k = \km(\phib)$ along $z_-(t)$ where
$dz/dt=\om_{0,k}$, which, when combined with the modulation equations
\eqref{eq:reducedwhitham}, results in the ODE
\begin{equation}
  \frac{d\km}{d\phib} = \frac{\omega_{0,\phib}}{2\phib-\omega_{0,\km}}.
\end{equation}
Substituting the linear dispersion relation \eqref{eq:lineardispreln}
into this equation and integrating yields an expression for $\km$ in
terms of the wave mean $\phib$ and an integration constant $\lambda$
\begin{equation}\label{eq:kminus}
  \km(\phib;\lambda)^2 = \frac{1}{2}\left( \lambda - \frac{2}{\phib} +
    \sqrt{\frac{\lambda}{\phib}(4+\phib\lambda)} \right).
\end{equation}
Matching to equation \eqref{eq:hopf} at the disturbance's initial
termini $z\in\{-w,0\}$ via the first of equation \eqref{eq:wavetrainEdges},
$k_-(\phib=1;\lambda) = 0$.  Then $\lambda = 1/2$.  This choice of
integration parameter results in the same expression for $k_-$ as
found at the DSW's harmonic edge from DSW fitting theory; see
equation \eqref{eq:DSWTrailingWavenum}.  
It will be useful in the next section to use the fact that the simple
wave curve $k = k_-(\phib;1/2)$ corresponds to the level curve
$\lambda = 1/2$ of the surface
\begin{equation}
  \label{eq:6}
  \lambda(k,\phib) = \frac{(1+\phib k^2)^2}{\phib (2 + \phib k^2)},
\end{equation}
obtained by inverting the relationship in equation \eqref{eq:kminus}.

The simple wave relationship $k = \km(\phib;1/2)$ in
equation \eqref{eq:kminus} provides the needed translation between the
initial condition for the mean $\phib(z,0) = 1+a_0(z)$ and the initial
condition for the wavenumber $k(z,0) = \km(1+a_0(z);1/2)$.  Then the
number of solitary waves is obtained from equation \eqref{eq:Nsoli} as
\begin{equation}
  \label{eq:NsoliFinal}
  \begin{split}
    \Ncal &= \frac{1}{2\pi}\intinf k(z,0)\rmd z
    = \frac{1}{2\pi} \intinf \km\left(1+a_0(z);\frac{1}{2}\right)\rmd
    z.
  \end{split}
\end{equation}
For the case when $a_0(z)$ is a box of width $w$ and height $a_m$
above a background of $1$, equation \eqref{eq:NsoliFinal} can be integrated exactly
\begin{equation}\label{eq:Nsoli_box}
  \Ncal = \frac{w}{4\pi} \sqrt{\frac{a_m-3}{1+a_m}+
    \sqrt{\frac{9+a_m}{1+a_m}}}.
\end{equation}
The small $a_m$ expansion of equation \eqref{eq:Nsoli_box} is
\begin{equation}
  \Ncal =  \frac{w\sqrt{a_m}}{\pi\sqrt{6}} + \Order{wa_m^{3/2}} ,
  \quad a_m \to 0 ,
\end{equation}
which agrees to leading order with the small-amplitude KdV result in
equation \eqref{eq:1} when we identify $u_m = a_m$.  The large $a_m$
approximation, on the other hand, is independent of box height to a
good approximation
\begin{equation}\label{eq:Nsoli_large_am}
  \Ncal = \frac{w\sqrt{2}}{4\pi} + \Order{\frac{w}{a_m^2}} , \quad
  a_m \to \infty .
\end{equation}

To compute the number of solitary waves for more general initial
profiles $a_0(z)$, equation \eqref{eq:NsoliFinal} can be integrated.
Of course, the number of solitary waves should be an integer whereas
$\Ncal$ continuously depends on the initial profile $a_0(z)$.  The
result is asymptotic, i.e., equation \eqref{eq:NsoliFinal} is
asymptotic to the number of solitary waves due to solitary wave
fission if $\Ncal \gg 1$.  Then, computing the ceiling, floor, or
rounding $\Ncal$ to the nearest integer are all asymptotically
equivalent.  If we approximate the initial disturbance by a box of
width $w$ and height $a_m$, equation \eqref{eq:Nsoli_box} gives an
explicit determination of when modulation theory for solitary wave
fission is valid, i.e., when the initial disturbance is sufficiently
wide.

\section{Distribution of solitary wave amplitudes}
\label{sec:distr-solit-wave}

Next, we seek an estimate for the amplitudes of the fissioned solitary
wavetrain.  Because the conduit solitary wave speed-amplitude relation
\eqref{eq:2} is monotonically increasing with amplitude, sufficiently
long evolution is expected to lead to the waves separating into an
amplitude-ordered train that are well isolated from one another.  We
will treat them as a non-interacting solitary wavetrain, a concept
that was recently exploited in \citep{maiden_hydrodynamic_2018} to
describe solitary wave interaction with a mean flow.  Here, we analyse
both the $\Acal \to 0$ (harmonic) and $k \to 0$ (solitary wave) limits
and identify a relationship between them.  This enables a mapping of
the initial profile to the long-time solitary wave amplitude
distribution.

\subsection{Harmonic limit}
\label{sec:harmonic-limit}

In the harmonic limit, $\Acal \to 0$ and
Equations \eqref{eq:reducedwhitham} and \eqref{eq:5} hold.  To compute the
total number of solitary waves in the previous section, we identified
the edges of the initial disturbance's support and set $k = 0$ at the
edges.  This calculation resulted in the simple wave relationship
determined by the level curve $\lambda(k,\phib) = 1/2$
(cf.~equation \eqref{eq:6}).  We now extend this to the
interior of the initial disturbance's support and study other level
curves, $\lambda(k,\phib) = $ constant, to identify the number of
solitary waves contained within a portion of the initial
disturbance.  
By solving for $\lambda$ when $k = k_-(1+a_0(z);\lambda) = 0$, we
therefore consider the level curves
$\lambda(k,\phib)=[2(1+a_0(z))]^{-1}\in[1/(2(1+a_m)),1/2]$.
        
We can now extend the calculation of the total number of solitary
waves $\Ncal$ to the number of solitary waves that emerge from the
section of the initial profile of total amplitude of at least
$\pbmin$. We modify equation \eqref{eq:NsoliFinal} for the total number of
fissioned solitary waves to integrate only over the initial profile
section in which $1 + a_0(z) \ge \pbmin$ (see
figure~\ref{fig:truncate_effect}(a)) and consider the $\lambda$-level
curve $\lambda(k,\phib)=1/(2\pbmin)$, determined by the zero
wavenumber condition $k_-(\pbmin;\lambda)=0$.  Then the number of
solitary waves for this truncated portion of the initial profile is
\begin{equation}
  \label{eq:NsoliLC}
  \begin{split}
    G(\pbmin) &= \frac{1}{2\pi} \int_{z_1(\pbmin)}^{z_2(\pbmin)}
    \km\left(1+a_0(z);\frac{1}{2\pbmin}\right) \rmd z, \quad
    \mathrm{for} ~ \pbmin \in [1,1+a_m]\\
    z_1\leq z_2 &\text{ such that } 1+a_0(z_{1,2})=\pbmin.
  \end{split}
\end{equation}
The justification for this calculation comes from hyperbolicity of the
modulation system \eqref{eq:whitham} in the requisite domain of
dependent variables \citep{maiden_modulations_2016} and the fact that
asymptotically, as $t \to \infty$, the region of influence of the
support of the $\lambda$ section of the initial profile is confined by
the modulation characteristics emanating from $z_2$ and the maximum
point $z_m$: $a_0(z_m)=a_m$---see \cite{el_asymptotic_2008}.

The goal now is to relate $\pbmin$ to the solitary wave amplitude
$\mathcal{A}$.  We will use the intermediate variable $\lambda$ to
relate the two.  With a slight abuse of notation, we define
$z_1(\lambda)$, $z_2(\lambda)$ as the $z$-values at which
$\phib = 1/(2\lambda)$ and $G(\lambda)$ as the number of
solitary waves that emerge from the $\lambda$-section of the initial
profile of total amplitude at least $1/(2\lambda)$
\begin{equation}\label{eq:Flambda}
  \begin{split}
    G(\lambda) &= \frac{1}{2\pi}
    \int_{z_1(\lambda)}^{z_2(\lambda)} \km(1+a_0(z);\lambda) \rmd z,
    \quad \mathrm{for}~ \lambda \in
    \left [ \frac{1}{2(1+a_m)},\frac{1}{2} \right ] ,\\
    z_1\leq z_2 &\text{ such that } 1+a_0(z_{1,2})=1/(2\lambda).
  \end{split}
\end{equation}
Since $G(\lambda)$ is an increasing function of $\lambda$ and its
maximum is the total number of solitary waves $\mathcal{N} = G(1/2)$
(cf.~equation \eqref{eq:NsoliFinal}), we define the normalised
cumulative density function (cdf) $\mathcal{G}(\lambda)$ as
\begin{equation}
  \label{eq:11}
  \mathcal{G}(\lambda) = 
  \frac{G(\lambda)}{\mathcal{N}} \in [0,1], \quad
  \mathrm{for} ~ \lambda \in
  \left [ \frac{1}{2(1+a_m)},\frac{1}{2} \right ] .
\end{equation}
Reconsidering the smoothed box and its numerical evolution from figure
\ref{fig:conduitIVP} in this way, profile sections for different
values of $\lambda$ are shown in figure \ref{fig:truncate_effect}(a)
and the expected solitary waves from each truncation are shown in
figure \ref{fig:truncate_effect}(b).
\begin{figure}
  \centering
  \subfigure[]{\includegraphics{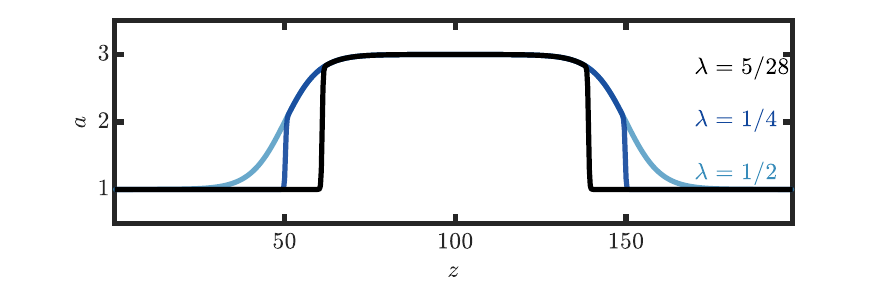}}
  \subfigure[]{\includegraphics{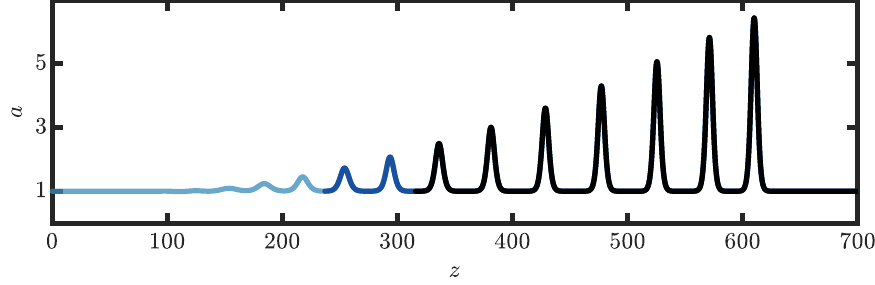}}
  \caption{(a) Sections of the initial profile from figure
    \ref{fig:conduitIVP} for different values of $\lambda$. The
    different $\lambda$ sections are identified by color and
    shading. (b) Contribution of each $\lambda$ section in terms of
    the produced solitary waves.}
  \label{fig:truncate_effect}
\end{figure}
The integral endpoints $z_1$ and $z_2$ for the initial condition shown
in figure~\ref{fig:truncate_effect}(a) are shown as a function of
$\lambda$ in figure \ref{fig:z_vs_l}.  The endpoints depend
monotonically on $\lambda$.
\begin{figure}
  \centering
  \includegraphics{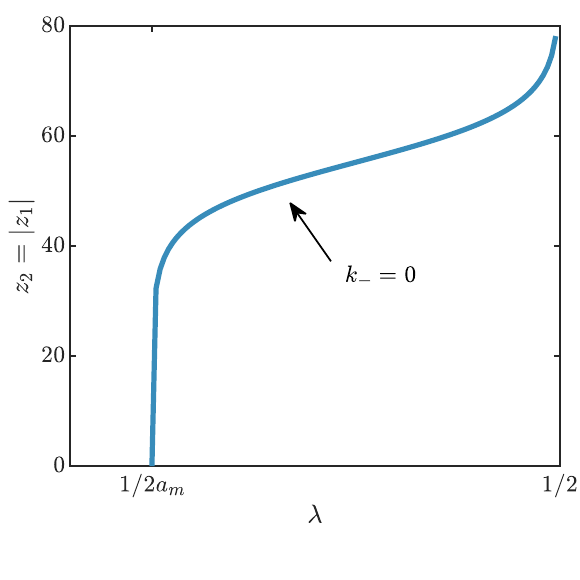} 
  \caption{Integral endpoints $z_{1,2}$ as a function of the
    integration constant, $\lambda$, for the profile in figure
    \ref{fig:truncate_effect}(a).}
  \label{fig:z_vs_l}
\end{figure}

\subsection{Solitary wave limit}
\label{sec:solitary-wave-limit}

So far, we have been focused on the number of fissioned solitary waves
emerging from a $\lambda$ section of the initial profile.  We need to
relate $\lambda$ to the amplitudes of the fissioned solitary waves.
For this, we now perform an analysis of the solitary wave limit
$k \to 0$ of the conduit-Whitham equations \eqref{eq:whitham}, which
describe the modulations of a non-interacting solitary wavetrain.  By
use of a clever change of variables
\citep{el_resolution_2005, el_dispersive_2016}, this limit can be put in
a form that is analogous to the harmonic limit analysis of equation
\eqref{eq:reducedwhitham}.  In particular, we will determine a
relationship between the wave amplitude and mean,
$\Acal=\Acal(\phib)$, that is valid along the characteristic family
associated with the propagation of non-interacting solitary waves.
This relationship will be a simple wave curve.

We now consider the conduit-Whitham equations \eqref{eq:whitham} in
the solitary wave limit $k\to0$.  Here, the wavelength $2\pi/k$ tends
to infinity, so again the contribution of oscillations is negligible
and averaging commutes $\bar{F(\phi)}=F(\phib)$
\citep{el_dispersive_2016}.  Then the modulation equations reduce to
the dispersionless mean flow equation and an equation for the solitary
wave amplitude $\Acal$ \citep{maiden_hydrodynamic_2018}
\begin{align}
  &k=0: \nonumber \\
  \label{eq:reducedwhithamA}
  & \phib_t + 2\phib\, \phib_z = 0, \\
  \label{eq:7}
  & \Acal_t + c_s(\phib+\Acal,\phib)\Acal_z + g(\Acal,\phib)\phib_z = 0, 
\end{align}
where $c_s$ is the solitary wave speed-amplitude relation
\eqref{eq:conduitsoliampl} and $g$ is a coupling function that we will
not need to explicitly determine.  We now introduce the convenient
change of modulation variables
$(\phib,\Acal,k) \to (\phib,\kt,\Lambda)$ \citep{el_resolution_2005}
\begin{equation}
  \kt = \pi\left( 
    \int_{\phi_1}^{\phi_2}\frac{\rmd\phi}{\sqrt{-g(\phi)}} 
  \right)^{-1}, \quad \Lambda = \frac{k}{\kt},
\end{equation}
where $\phi_{1,2}$ are the two smaller roots of the right side of the
periodic wave ODE \eqref{eq:conduitPeriodic}.  

This change of variable is based on the idea of a conjugate conduit
equation, where $\At(\zt,\Tt) = a(i\zt,i\Tt)$ is substituted into the
conduit equation \eqref{eq:conduit} so that it becomes
\begin{equation}
  \label{eq:conjconduit}
  \At_{\Tt} + \left (\At^2 \right)_{\zt} + \left( \At^2 \left( \At^{-1}\At_{\Tt}
    \right)_{\zt} \right)_{\zt} = 0.
\end{equation}
The parameter $\kt$ is the wavenumber of the conjugate travelling wave
that satisfies the ODE
\begin{equation}
  \label{eq:8}
  \begin{array}{ccc}
    \left(\phit_{\tht}\right)=-g(\phit),
    & \phit(\tht+2\pi) =
      \phit(\tht)  &  \tht = \kt \zt - \wt \Tt,
  \end{array}
\end{equation}
with the conjugate linear dispersion relation
\begin{equation}\label{eq:conjlineardisp}
  \wt_0(\kt,\phib) = \frac{2\kt\phib}{1-\kt^2\phib }. 
\end{equation}
We require that periodic solutions $\phi(\theta)$ and $\phit(\tht)$ to
equations \eqref{eq:conduitPeriodic} and \eqref{eq:8}, respectively,
have identical phase velocities $c_p=\om/k=\wt/\kt$, thus
$\om=\Lambda\wt$.  The benefit of this formulation is that the
solitary wave limit of the conduit equation periodic wave is the
harmonic limit of the conjugate conduit equation periodic wave, and
can be leveraged as such.  It allows for a formulation of the solitary
wave limit that is symmetric to the harmonic limit. By substituting
$k=\Lambda\kt$, $\om=\Lambda\wt$ into the equation for conservation of
waves $k_t+\om_z=0$, we obtain.
\begin{equation}\label{eq:ConsWaveskt}
  \kt\Lambda_t + \wt\Lambda_z + 
  \Lambda\left( \kt_t + \wt_z \right) = 0.
\end{equation}
In the solitary wave limit, $k\to 0$ and therefore $\Lambda\to0$, but
this limit is a singular one in that $|k_x|,|k_t| \to \infty$ for a
DSW and therefore $|\Lambda_t|,|\Lambda_x| \to \infty$
\citep{el_resolution_2005}.  We therefore consider equation
\eqref{eq:ConsWaveskt} when $|\Lambda|\ll|\Lambda_t|,|\Lambda_x|$ to
obtain the leading order equation
\begin{equation}\label{eq:LambdaLO}
  \Lambda_t + \frac{\wt_0}{\kt}\Lambda_z = 0.
\end{equation}
This equation admits the characteristics 
\begin{equation}
  \frac{\dop{z}}{\dop{t}}=\frac{\wt_0(\kt,\phib)}{\kt}=c_p.
\end{equation}
The specific characteristic in which $\Lambda=0$ corresponds to $k=0$ and is
the solitary wave edge of the wavetrain.  Along $\Lambda=0$, the
characteristic speed $c_p$ is equal to the solitary wave
speed-amplitude relation $c_s$ \eqref{eq:conduitsoliampl}.  This
relation $c_p=\wt(\kt,\phib)/\kt=c_s(\phib+\Acal,\phib)$ determines the change of
variable $(\phib,\Acal) \to (\phib,\kt)$ when $\Lambda = 0$ for a
non-interacting solitary wavetrain
\begin{equation}\label{eq:kt2ampl}
  \kt^2 = \frac{1}{\phib}-\frac{2}{c_s(\phib+\Acal,\phib)}.
\end{equation}
One can see using \eqref{eq:conduitsoliampl} that $\Acal \to 0$
implies $\kt \to 0$ and vice versa so $\kt$ is an amplitude type
variable \citep{el_resolution_2005, lowman_dispersive_2013}.

The next order equation when $|\Lambda|\ll|\Lambda_t|,|\Lambda_x|$ is
\begin{equation}
  \label{eq:9}
  \kt_t +\left(\wt_0\right)_z = 0,\quad \text{on} \quad
  \frac{\dop{z}}{\dop{t}}=\frac{\wt_0(\kt,\phib)}{\kt}. 
\end{equation}
Similar to $k_-$ for harmonic waves, the simple wave assumption
$\kt = \kt_+(\phib)$ results in the ordinary differential equation
\begin{equation}
  \label{eq:17}
  \frac{\dop{\kt_+}}{\dop{\phib}} = \frac{\wt_{0,\phib}}{2\phib-\wt_{0,\kt}},
\end{equation}
whose integration results in
\begin{equation}\label{eq:ktsq}
  \ktp(\phib,\lt)^2 = \frac{1}{2}\left(-\lt +
    \frac{2}{\phib} - \sqrt{\frac{\lt}{\phib}(4+\lt\phib)}\right),
\end{equation}
where $\lt$ is the integration constant.

\subsection{Combined solitary wave and harmonic limits}
\label{sec:comb-solit-wave}

Combining the simple wave results for both the harmonic wave limit
$\Acal \to 0$ and the solitary wave limit $k \to 0$, we have the
following characteristic integrals
\begin{align}
  \label{eq:systemkm}
  I_H = \Bigg \{ \Acal=0,\quad \km^2 = \frac{1}{2}\left( \lambda -
  \frac{2}{\phib} + \sqrt{\frac{\lambda}{\phib}(4+\phib\lambda)}
  \right)\Bigg \},
  & \text{ on } \frac{\dop{z}}{\dop{t}}=\omega_{0,k}(k,\phib) ,
  \\
  \label{eq:systemkt}
  I_S = \Bigg\{ k=0,\quad \ktp^2 = \frac{1}{2}\left(- \lt +
  \frac{2}{\phib} - \sqrt{\frac{\lt}{\phib}(4+\lt\phib)}\right) \Bigg \}  , &  \text{ on
  }\frac{\dop{z}}{\dop{t}}=\frac{\wt_0(\kt,\phib)}{\kt}. 
\end{align}
Compatibility between the harmonic and the solitary wave regimes
within a single structure---the DSW---implies a relation between the
integration constants $\lambda$ and $\lt$. Indeed, if $\phib$ is such
that $\km(\phib;\lambda)=0$ in $I_H$ then, simultaneously,
$\ktp(\phib;\lt)=0$ in $I_S$ (see \citep{el_asymptotic_2008} for
details).  By eliminating $\phib$, we obtain
\begin{equation}
  \label{eq:12}
  \lt=\lambda .
\end{equation}
We remark that this same result---equivalence of the integral curve
parameters $\lambda = \lt$ for the harmonic and solitary wave
reductions---was obtained for both the KdV and Serre equations in
\cite{el_asymptotic_2008}.
        
In long time, the solitary waves are travelling on a unit background,
so inserting $\phib=1$ and $\lt=\lambda$ into equation
\eqref{eq:ktsq} relates $\lambda$ to $\kt$
\begin{equation}\label{eq:lambdakt}
  \lambda = \frac{(\kt^2-1)^2}{\kt^2 - 2}. 
\end{equation}
Then equations \eqref{eq:kt2ampl} and \eqref{eq:lambdakt} together
identify the desired relationship between $\lambda$ and $\Acal$, the
solitary wave amplitude measured from unit background $\phib = 1$
\begin{equation}
  \label{eq:lambdaAcal}
  \lambda(\Acal) = \frac{4}{c_s(1+\Acal,1)^2+2c_s(1+\Acal,1)} .
\end{equation}
Since $\lambda \in [1/(2(1+a_m)),1/2]$, $\mathcal{A}$ is limited to
the values $[0,\mathcal{A}_{\rm max}]$, where $\lambda(0) = 1/2$ and
$\mathcal{A}_{\rm max}$ is defined such that
\begin{equation}
  \label{eq:13}
  \lambda(\mathcal{A}_{\rm max}) = \frac{1}{2(1+a_m)} ,
\end{equation}
thus $\lambda$ is a decreasing function of $\Acal$.  Using
\eqref{eq:lambdaAcal}, we obtain the implicit expression for
$\Acal_{\rm max}$
\begin{equation}
  \label{eq:18}
  c_s(1+\Acal_{\rm max},1) = \sqrt{9 + 8 a_m} - 1 .
\end{equation}
This equation for the total amplitude $1+\Acal_{\rm max}$ is the same
expression one obtains for the DSW's leading edge solitary wave
amplitude $a_+$ in equation \eqref{eq:22} that results from an initial
jump of height $a_m$.  This concurs with our interpretation of the
initial box evolution as the generation of a DSW on the right and a RW
on the left.  Moreover, being entirely determined by the box height,
the lead solitary wave's amplitude is predicted to be independent of
box width.

Then $G(\lambda)$ from equation (\ref{eq:Flambda}) can be written in
terms of $\Acal$
\begin{equation}\label{eq:FlambdaAcal}
  F(\Acal) = G(\lambda(\mathcal{A})) = \frac{1}{2\pi}
  \int_{z_1(\lambda(\Acal))}^{z_2(\lambda(\Acal))}
  \km(1+a_0(z);\lambda(\Acal)) \rmd z, \quad \mathcal{A} \in
  [0,\mathcal{A}_{\rm max}] .
\end{equation}
Because $\lambda(\Acal)$ is a decreasing function of $\Acal$ and
$G(\lambda)$ is an increasing function of $\lambda$, the normalised
cdf of the fissioned solitary wave amplitude distribution is
\begin{align}
  \label{eq:amplcdf}
  \mathcal{F}(\Acal) &= 1 - \frac{F(\mathcal{A})}{\mathcal{N}} = 1 -
                       \mathcal{G}(\lambda(\mathcal{A})), \quad 
                       \mathcal{A} \in [0,\mathcal{A}_{\rm max}].
\end{align}
Since this distribution is continuous and we have a fixed number of
solitary waves, we will use the quantiled discretization of this
distribution for comparison with experiment and numerics.
        
We now attempt to explain what is seen in figure
\ref{fig:boxsolireln_w}(b), namely that initial conditions of
differing widths but otherwise the same height have approximately the
same normalised cdf.  To do so, we approximate the initial condition
with a box of width $w$ and height $a_m$.  Thus the normalised cdf in
$\lambda$ is
\begin{align}
  \mathcal{G}(\lambda) = \frac{\int_{-w}^0 k_-(1+a_m;\lambda)\,
  \mathrm{d}z}{\mathcal{N}}, \quad \lambda \in \left [
    \frac{1}{2(1+a_m)}, \frac{1}{2} \right ].
\end{align}
Since there is no variation in $z$, the numerator can be trivially
integrated
\begin{equation}
  \begin{split}
    \mathcal{G}(\lambda) &= \frac{w}{2\pi}\sqrt{\frac{1}{2}\left(
        \lambda -  \frac{2}{1+a_m} +
        \sqrt{\frac{\lambda}{1+a_m}(4+(1+a_m)\lambda)}
      \right)}/\mathcal{N}, \\
    \lambda &\in \left [ \frac{1}{2(1+a_m)}, \frac{1}{2} \right ].
  \end{split}
\end{equation}
Then, inserting $\mathcal{N}$ from equation \eqref{eq:Nsoli_box} leads to
no $w$-dependence in the normalised cdf
\begin{equation}\label{eq:Iw_norm}
  \mathcal{G}(\lambda) = \sqrt{ \frac{ 2\lambda a_m +2\lambda-4 +
      2\sqrt{\lambda(1+a_m)(4+(1+a_m)\lambda)}}
    {a_m-3+\sqrt{(9+a_m)(1+a_m)}}}, \quad \lambda \in \left [
    \frac{1}{2(1+a_m)}, \frac{1}{2} \right ].
\end{equation}
This approximation is valid as long as the edges of the disturbance
transition over a small $z$ relative to $w$.

Then the normalised cdf of the amplitude distribution is obtained from
\eqref{eq:Iw_norm} by substituting the functional relationship
$\lambda(\Acal)$ from equation \eqref{eq:lambdaAcal} and noting the
reflection \eqref{eq:amplcdf}
\begin{equation}\label{eq:Iw_a}
  \begin{split}
    \mathcal{F}(\mathcal{A}) &= 1 - \sqrt{ \frac{ 2\lambda(\Acal) a_m +
        2\lambda(\Acal) - 4 +
        2\sqrt{\lambda(\Acal)(1+a_m)(4+(1+a_m) \lambda(\Acal))}}   {a_m-
        3+ \sqrt{(9+a_m) (1+a_m)}}}, \\
    \mathcal{A} &\in [0,\mathcal{A}_{\rm max}] .
  \end{split}
\end{equation}
An asymptotic expansion of $\mathcal{F}$ in \eqref{eq:Iw_a} for small
$\Acal$ and $a_m$ yields
\begin{equation}
  \label{eq:21}
  \mathcal{F}(\Acal) \sim 1 - \sqrt{1 - \frac{\Acal}{2 a_m}}, \quad
  \Acal \in [0,2 a_m], \quad 0 < a_m \ll 1, 
\end{equation}
which agrees with the weakly nonlinear KdV result \eqref{eq:1}.

\subsection{Summary of the solitary wave resolution method}
\label{sec:general-method}

The above derivation is readily generalised.  Consider the initial
value problem for a general dispersive hydrodynamic equation
\begin{equation}
  \label{eq:16}
  \begin{split}
    u_t + V(u) u_x = D[u]_x, \quad x \in \mathbb{R}, \quad t > 0,\\
    u(x,0) = u_0(x), \quad \lim_{|x| \to \infty} u_0(x) = u_\infty,
  \end{split}
\end{equation}
with integro-differential operator $D$ yielding the real valued,
linear dispersion relation $\om_0(k,\ub)$ with negative dispersion
$\om_{0,kk} < 0$. Let equation \eqref{eq:16} support solitary wave
solutions propagating on the background $\ub$ and characterised by the
speed-amplitude relation $c_s(\ub+\mathcal{A}, \ub)$, where
$\mathcal{A}$ is the soliton amplitude measured from background. Now
we introduce $\km(\ub;\lambda)$ as the solution of the ODE
\begin{equation}
  \frac{d \km}{d\ub} =
  \left[\frac{\omega_{0,\ub}}{V(\ub)-\omega_{0,k}}\right]_{k=\km},  
\end{equation}
with $\lambda$ a constant of integration.  The number of solitary
waves resulting from the temporal evolution of $u_0(x)$ can then be
calculated as
\begin{equation}
  \Ncal = \frac{1}{2\pi}\intinf k_-\left( u_0(x); \lambda_\infty \right)\dop{x},
\end{equation}
where $\lambda_\infty$ is obtained from the boundary condition
$\km(u_\infty; \lambda_\infty)=0$.  For $u_0(x)$ in the form of a box
of width $w$ and height $u_m$,
\begin{equation}
  \label{eq:14}
  \Ncal = \frac{w}{2\pi}k_-(u_m; \lambda_\infty).
\end{equation}

For the solitary wave amplitudes, we have the generic formula in terms
of the relationship between the integration constant $\lambda$ and the
cutoff mean $\ub$
\begin{equation}
  \label{eq:15}
  \begin{split}
    \mathcal{G}(\lambda) = \frac{1}{2\pi\mathcal{N}}
    \int_{x_1(\lambda)}^{x_2(\lambda)} \km(u_0(x);\lambda) \rmd x,
    \quad \lambda \in [\lambda_m,\lambda_\infty]\\ 
    x_1(\lambda) \leq x_2(\lambda) \text{ such that }
    u_0(x_{1,2}(\lambda))=\ub(\lambda),
  \end{split}
\end{equation}
where $\lambda_m$ is defined by $\km(u_m;\lambda_m) = 0$.  Here, we
are assuming that $k_-(\ub;\lambda)$, hence $\mathcal{G}(\lambda)$, is
an increasing function of $\lambda$.  Then to obtain
$\lambda=\lambda(\Acal)$, one first solves the ODE
\begin{equation}
  \label{eq:eq4d44}
  \frac{d \ktp}{d\ub} =
  \left[\frac{\wt_{0,\ub}}{V(\ub)-\wt_{0,\kt}}\right]_{\kt=\ktp}, 
\end{equation}
where $\wt_0(\kt,\ub)=-i\omega_0(i\kt,\ub)$.  The solution of equation
\eqref{eq:eq4d44} is $\ktp(\ub;\lt)$, where $\lt$ is a constant of
integration.  Setting $k(\ub;\lambda)=\kt(\ub;\lt)=0$ gives the
relationship between $\lambda$ and $\lt$.  Substituting
$\kt=\kt(\ub;\lt(\lambda))$ into $\wt/\kt=c_s(\ub+\Acal,\ub)$ yields
the desired $\lambda=\lambda(\Acal)$ and
$\mathcal{F}(\mathcal{A}) = 1-\mathcal{G}(\lambda(\mathcal{A}))$ for
$\mathcal{A} \in [0,\mathcal{A}_{\rm max}]$ where
$\mathcal{A}_{\rm max}$ satisfies
\begin{equation}
  \label{eq:23}
  \lambda(\Acal_{\rm max}) = \lambda_m .
\end{equation}
Here, we are assuming that $\lambda$ is a decreasing function of
$\Acal$.
        
Two of the main results of this paper do not depend on the system
under study so long as the necessary pre-requisites of the solitary
wave resolution method are satisfied.  Equation \eqref{eq:14} for a
pure box initial condition is always linear in the box width.  Also,
the maximum solitary wave amplitude $\Acal_{\rm max}$ and the
normalised cumulative amplitude distribution function for a box,
$\mathcal{F}(\Acal)$, are independent of box width.  These results can
be used, for example, to identify the initial box height that yields a
desired lead solitary wave with amplitude $\Acal_*$ by solving
$\lambda_* = \lambda(\Acal_*)$ where $k_-(u_*;\lambda_*) = 0$ for the
box height $u_*$ and the box width
$w_* = 2\pi \mathcal{N}_*/k_-(u_*;\lambda_\infty)$ that results in the
desired number of solitary waves $\mathcal{N}_*$.

\begin{figure}
  \centering
  \subfigure[]{\includegraphics{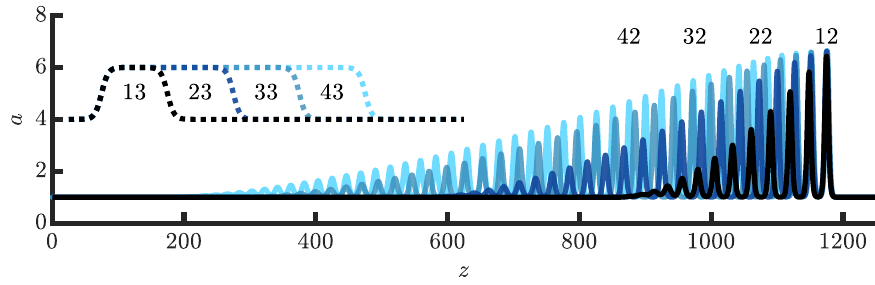}}
  \subfigure[]{\includegraphics{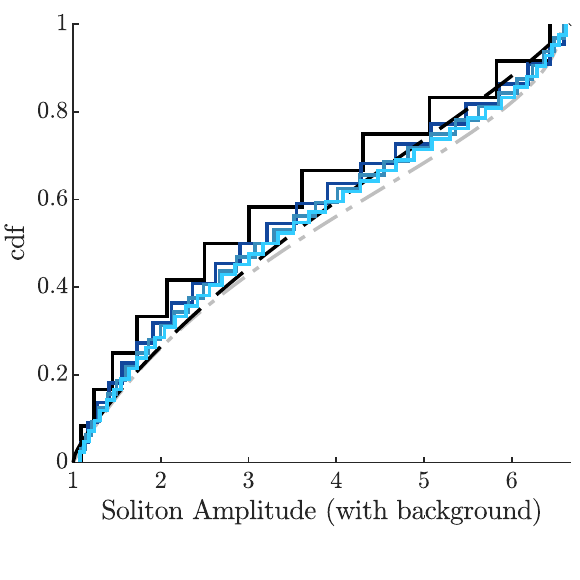}}
  \caption{(a) Numerical simulations of boxes of different widths and
    the ensuing solitary waves. Note the evolutions here are at
    different times and shifted to align so as to better illustrate
    the similarities and differences in the amplitude distributions.
    Integers reported above the solitary wavetrains are the number of
    observed solitary waves to be compared with the prediction
    \eqref{eq:NsoliFinal} reported inside the corresponding box
    initial condition. (b) (solid) Observed amplitude distributions
    from the same simulations. (dashed) Predicted amplitude
    distribution from equation \eqref{eq:amplcdf}. (dash-dotted)
    Predicted amplitude distribution \eqref{eq:Iw_a} for a box.}
  \label{fig:boxsolireln_w}
\end{figure}

\subsection{Numerical methods}
Direct numerical simulations of the conduit equation were undertaken
following the method described in \citep{maiden_modulations_2016}.
Equation \eqref{eq:conduit} is rewritten as two coupled equations, the
spatial discretisation utilises fourth-order finite differences with
periodic boundary conditions, and the temporal evolution is via a
medium-order Runge-Kutta method.  Numerical results presented show how
the long-time box evolution is altered by width in figure
\ref{fig:boxsolireln_w} and by height in figure
\ref{fig:boxsolireln_a}.  We observe that the number of solitary waves
produced approximately changes linearly with the width but does not
change significantly with box height past a certain height.  We
observe that the amplitude distributions change with box height but
not with width.

\begin{figure}
  \centering
  \subfigure[]{\includegraphics{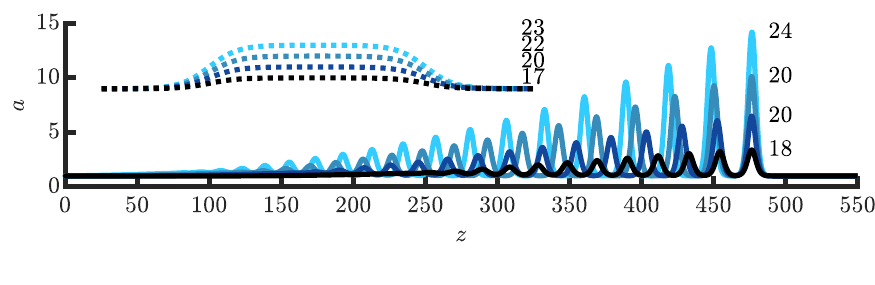}}
  \subfigure[]{\includegraphics{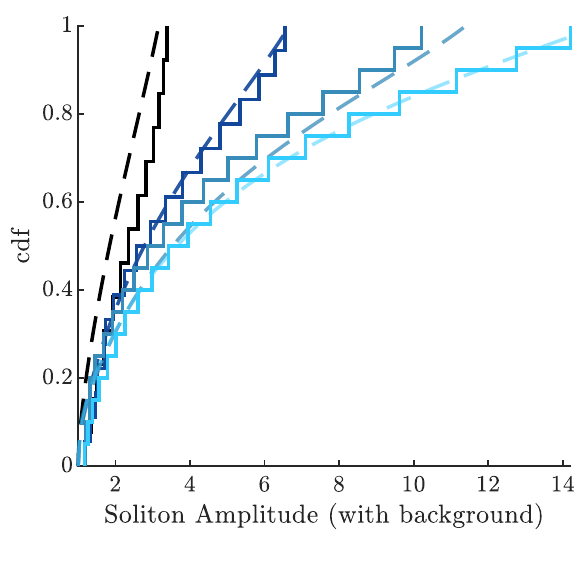}}
  \caption{(a) Numerical simulations of boxes of different heights and
    the ensuing solitary waves. The solitary wave trains are plotted
    at different times so as to better illustrate the similarities and
    differences in the amplitude distributions. The expected solitary
    wave counts (adjacent to the initial profile) and the observed
    number (adjacent to the solitary wave train) do not change much
    past a certain initial condition amplitude, as expected from
    equation \eqref{eq:Nsoli_large_am}. (b) (solid) Amplitude
    distributions from the same simulations. (dashed) Predicted
    amplitude distribution from equation \eqref{eq:amplcdf}.}
  \label{fig:boxsolireln_a}
\end{figure}

We also numerically investigate our basic hypothesis that an initial,
broad disturbance for the conduit equation results primarily in the
fission of solitary waves.  In order to quantify this, we consider one
simulation that represents an edge case in which a smoothed box with
amplitude $a_m = 0.88$ and width $w = 96$ results in a relatively
small number (9) of solitary waves and agrees with the predicted
number from equation~\eqref{eq:NsoliFinal}.  The initial and final (at
$t = 350$) profiles are shown in figure~\ref{fig:small_box}.  Solitary
waves travel faster than the long wave speed $c_s(a_s,1) > 2$, whereas
dispersive waves propagate with the group velocity that is slower and
exhibits a minimum $-1/4 \le \partial_k \omega_0(k,1) \le 2$. We
identify the dividing location in figure~\ref{fig:small_box} between
small amplitude dispersive waves and the solitary wavetrain as the
first $z$ value, $z_* = 1000$ here, in which the solitary wavetrain
departs from unity.  Over the entire domain $[0,L]$ ($L = 1500$ here)
and each of the two subintervals $[0,z_*]$, $[z_*,L]$, we compute
integrals of the conserved densities $a-1$, $1-1/a-a_z^2/a^2$ and the
nonnegative density $(a-1)^2$ at the final time.  The results are
reported in table~\ref{tab:wave_contributions}.  In all cases, the
small amplitude dispersive wave contributions are less than 1\% of the
total.  Consequently, the solitary wavetrain dominates these integral
quantities and our basic hypothesis is confirmed.
\begin{table}
  \centering
  \begin{tabular}{cccc}
    $I = $ & $[0,L]$ & $[0,z_*]$ & $[z_*,L]$ \\ \hline
    $\int_I (a-1)\, \mathrm{d}z$ & 84.48 & -0.34 & 84.82 \\[3mm]
    $\displaystyle \int_I \left ( 1-\frac{1}{a}-\frac{a_z^2}{a^2} \right)\,
    \mathrm{d}z$ & 46.36 & -0.35 & 46.71 \\[6mm]
    $\int_I (a-1)^2\, \mathrm{d}z$ & 81.39 & 0.01 & 81.38
  \end{tabular}
  \caption{Integrals of conserved and nonconserved quantities for the
    solution depicted in figure~\ref{fig:small_box} at $t = 350$ over
    three spatial intervals corresponding to the whole
    domain $[0,L]$, the subinterval containing only small amplitude,
    dispersive waves $[0,z_*]$, and the solitary wavetrain subinterval
    $[z_*,L]$ ($z_* = 1000$, $L = 1500$).}
  \label{tab:wave_contributions}
\end{table}

\begin{figure}
  \centering
  \includegraphics[scale=0.25]{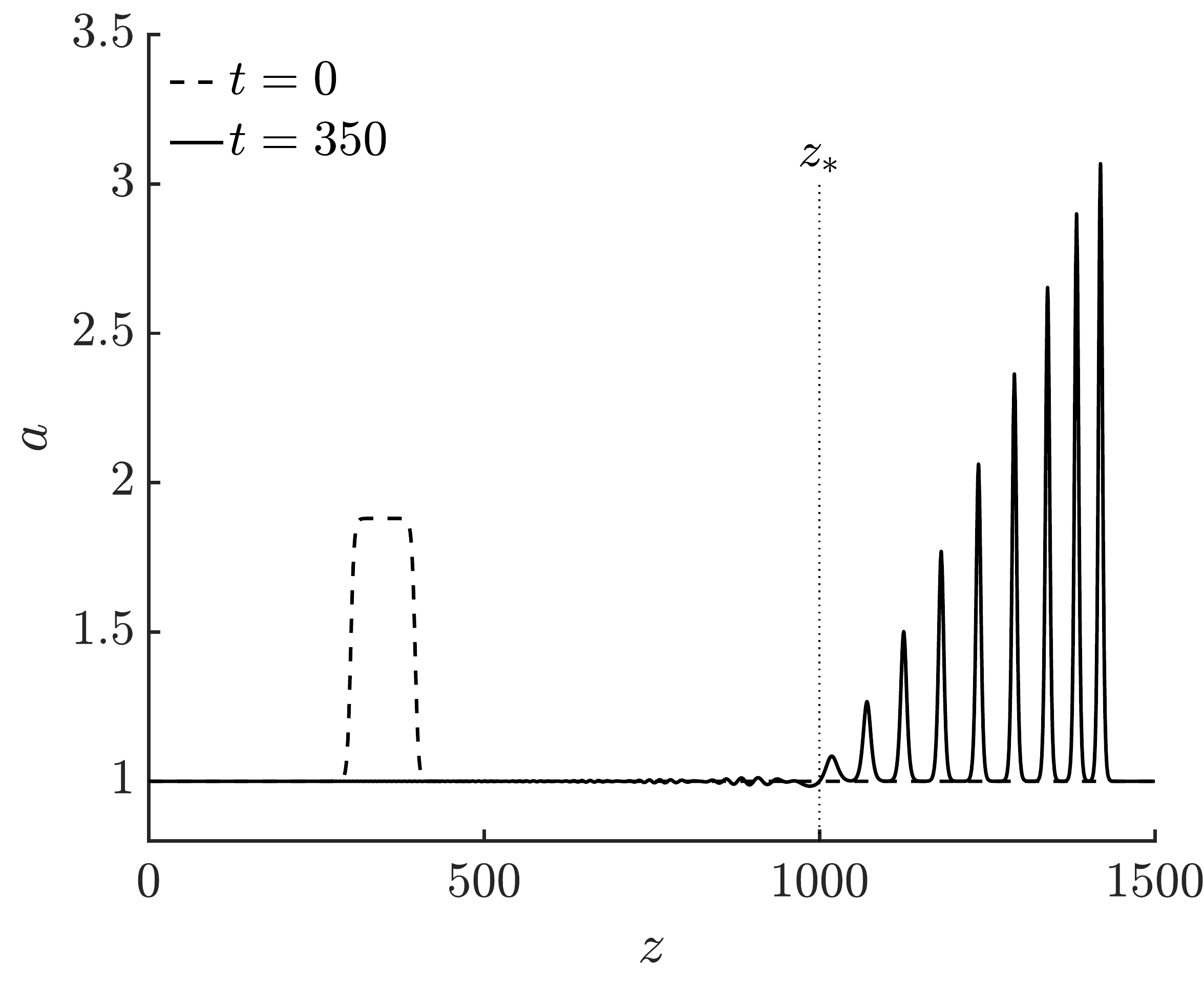}
  \caption{Initial (dashed) and final (solid) profiles for a numerical
    simulation of the conduit equation box problem ($a_m = 0.88$, $w =
    96$) that results in 9 solitary waves.  The location $z_* = 1000$
    separates the solitary wavetrain to the right from small amplitude
    dispersive radiation.}
  \label{fig:small_box}
\end{figure}

\section{Comparison of modulation theory with experiment}
\label{sec:comp-exper}          

Theory predictions for the number of solitary waves in physical
experiments are reported in figure \ref{fig:results_N}.  We calculate
the prediction $\Ncal$ from a smoothed version of the viscous fluid
conduit's observed profile at the time of wavebreaking. 
Filled dots and the vertical axis in figure \ref{fig:results_N}(a)
report the number of observed solitary waves for each trial.  We
observe excellent agreement between experiment and theory, with the
observed number of solitary waves being at most two away from the
predicted value.  Consequently, there is a decrease in the relative
percent error as the total number of solitary waves increases, as
shown in figure \ref{fig:results_N}(b).

    
\begin{figure}
  \centering
  \subfigure[]{\includegraphics{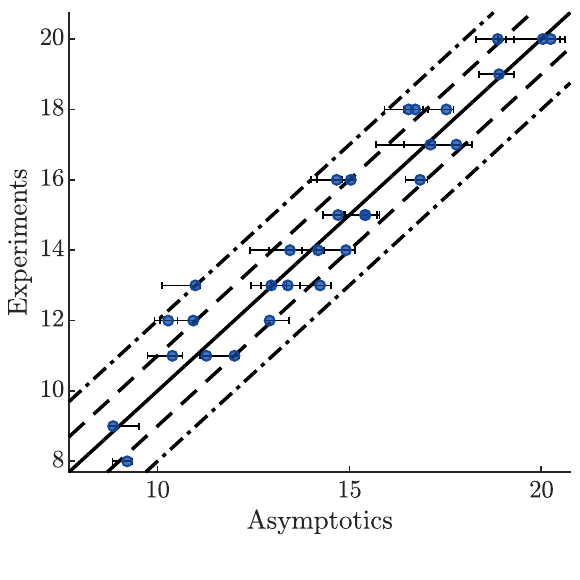}}
  \subfigure[]{\includegraphics{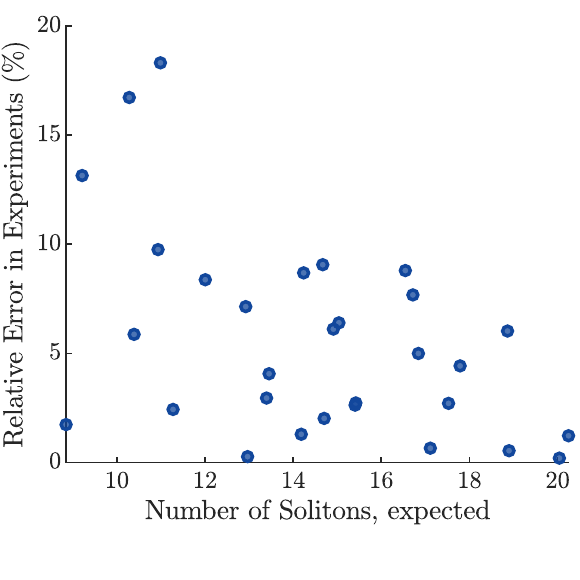}}
  \caption{ (a) Number of solitary waves from experiment (circles)
    versus the number expected from equation \eqref{eq:Nsoli}. The
    dashed and dash-dotted lines represent one and two solitary waves
    away from the expected 1:1 relationship.  (b) Percent relative
    error versus the expected number of solitary waves.}
  \label{fig:results_N}
\end{figure}
Figure \ref{fig:results_F} details comparisons between asymptotic
predictions and physical observations of the solitary wave amplitude
cumulative distribution functions.  The prediction from KdV analysis
is included. For the amplitude distribution $\mathcal{F}(\Acal)$, any
relative minimum in the initial profile results in unphysical
predictions.  Therefore, instead of using the true profile generated
from boundary control, we use an averaged version, similar to that
used in numerical experiments (see the inset of figure
\ref{fig:conduitIVP}).  We fit the box amplitude $a_m$ and width $w$
by extracting these values from the experimental time, location, and
mean height of breaking as determined by our previously introduced
inflection point criterion \citep{anderson_controlling_2019}.  The box
profile is approximated by the following formula
\begin{equation}
  \label{eq:10}
  a_0(z) = \frac{a_m}{2}\mathrm{tanh}\left ( \frac{z}{2.5} \right ) -
  \frac{a_m}{2} \mathrm{tanh} \left ( \frac{z-w}{2.5} \right ) ,
\end{equation}
where the tanh nondimensional width 2.5 was identified as a good fit
across all reported trials to both the leading edge transition and the
final amplitude distributions.  Although our analysis is explicit for
sharp box profiles, we find that smoothing the box does slightly
influence the smaller amplitude soliton distribution as depicted in
figure~\ref{fig:my_label} (see dashed versus dash-dotted curves).
\begin{figure}
  \centering
  \subfigure[]{\includegraphics{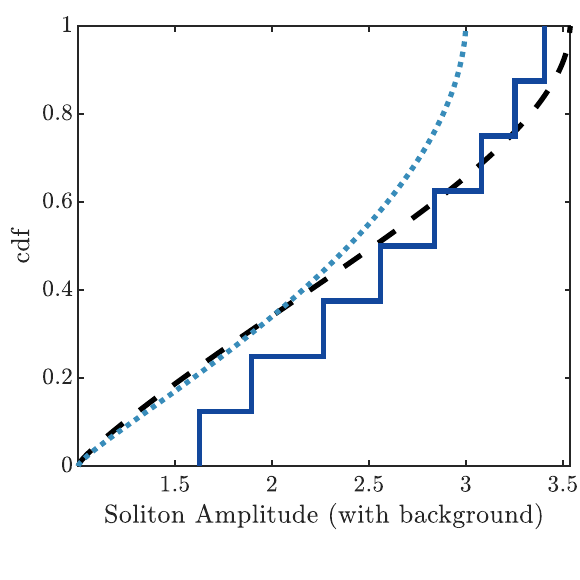}}
  \subfigure[]{\includegraphics{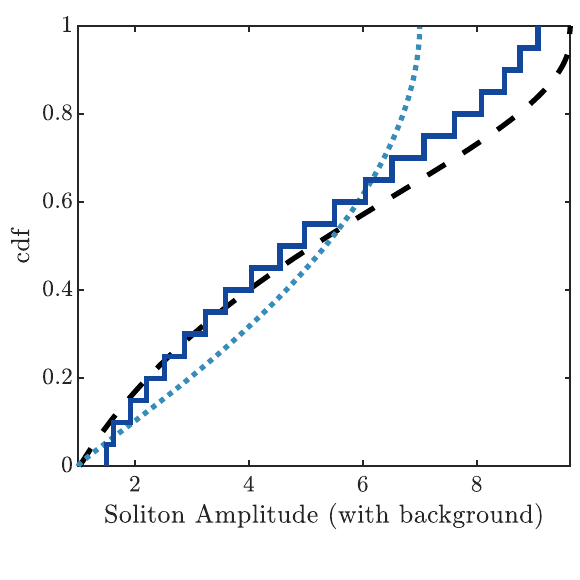}}
  \caption{Cdfs of amplitude distributions from selected experiments
    (solid line), with the asymptotic prediction from the conduit
    (dashed line) and KdV (dotted line) equations. Each step in the
    experimental cdf corresponds to a solitary wave. Box parameters:
    (a) width $=\SI{20}{\centi\meter}$ ($w = 90$), $a_m=2$ (b) width
    $=\SI{40}{\centi\meter}$ ($w = 178$), $a_m=4$}
  \label{fig:results_F}
\end{figure}
    
We also observe a change in conduit diameter of roughly $15\%$ from
the bottom of the apparatus to the location of solitary wave
data-taking.  While this does not affect $\Ncal$, this is observed to
have a profound affect on $\mathcal{F}(\Acal)$.  To compensate for
this, we use the amplitude prediction from
\citep{maiden_hydrodynamic_2018} for a solitary wave on a changing
background for the conduit equation.  We measure error via the
$\infty$-norm, as this relates to the Kolmogorov-Smirnov test for
comparing cdfs in statistics.  Across all trials, the conduit
prediction has roughly half the error as the prediction from the
rescaled KdV prediction.
    
We also compare our results to the explicit formula for a box
\eqref{eq:Iw_norm}. We observe in figure \ref{fig:my_label} that, as
the initial condition's width increases---corresponding to a larger
number of solitary waves therefore improving the asymptotic
approximation---the observed distribution approaches the expected
distribution that is independent of width.

\begin{figure}
  \centering
  \includegraphics{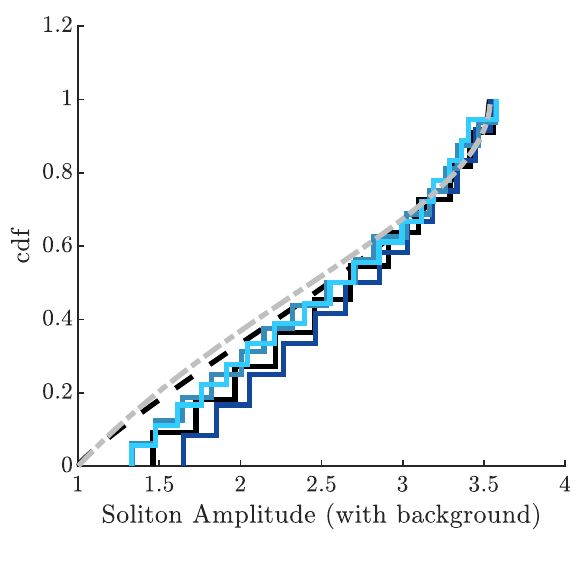}
  \caption{Normalised cumulative distribution functions for solitary
    wave amplitudes.  Experiments (stairs) compare favorably to the
    predictions from equation \eqref{eq:amplcdf} for the smoothed box
    (dashed) and predictions from equation \eqref{eq:Iw_a} for a pure
    box (dash-dotted). Color scale corresponds to initial conditions
    where $a_m=2$ and dimensional widths (light to dark)
    $25,30,35,\SI{40}{\centi\meter}$ corresponding to nondimensional
    widths $112,134,156,178$.}
  \label{fig:my_label}
\end{figure}

Our final comparison between experiment and theory involves the
spatio-temporal data set reported in figure~\ref{fig:expt_setup}(b).
Utilising the nominal, measured experimental parameter values reported
in the caption of that figure, we determine the length and time
scalings for the conduit equation in \eqref{eq:scaling1} to be
$R_0/\sqrt{8\epsilon} = \SI{1.6}{\milli\meter}$ and
$R_0/(U_0\sqrt{8\epsilon}) = \SI{1.17}{\second}$, respectively.  These
scalings and the measured parameters determine the initial,
nondimensional box width $w = 156$ and height $a_m = 1.6$.  A
numerical simulation of the conduit equation with this smoothed box
initial data (c.f.~equation~\eqref{eq:10}) and these scalings is shown in
figure~\ref{fig:expt_numerics_evolve}(a).  We report the nondimensional
diameter $\sqrt{a}$ in the figure (black curves) in order to directly
compare the numerics with experiment.  The conduit equation simulation
domain was taken to be larger ($\SI{160}{\centi\meter}$ in
\ref{fig:expt_numerics_evolve}(a) and $\SI{180}{\centi\meter}$ in
\ref{fig:expt_numerics_evolve}(b)) than the view shown so that a
portion of the initial box is outside the displayed viewing window.

\begin{figure}
  \centering
  \subfigure[Measured experimental
  parameters.]{\includegraphics{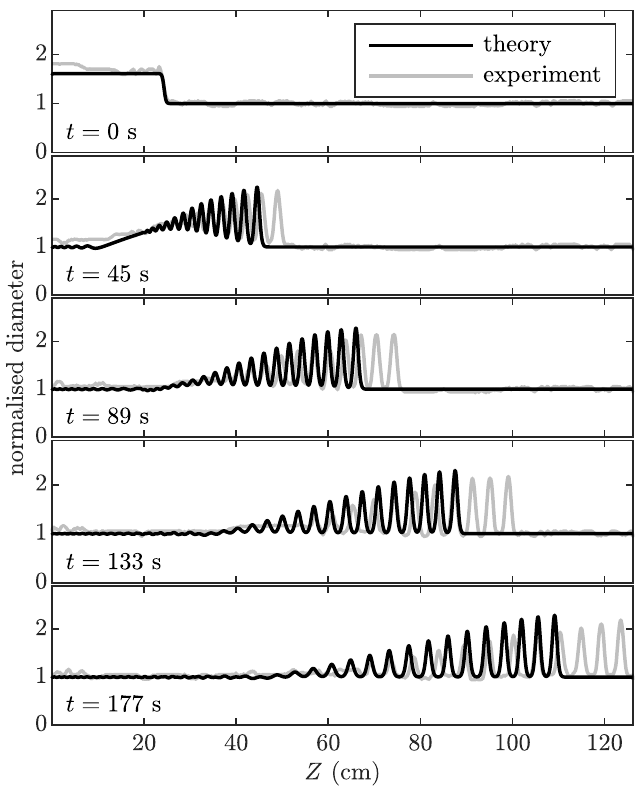}}
  \subfigure[Fitted experimental parameters.]{\includegraphics{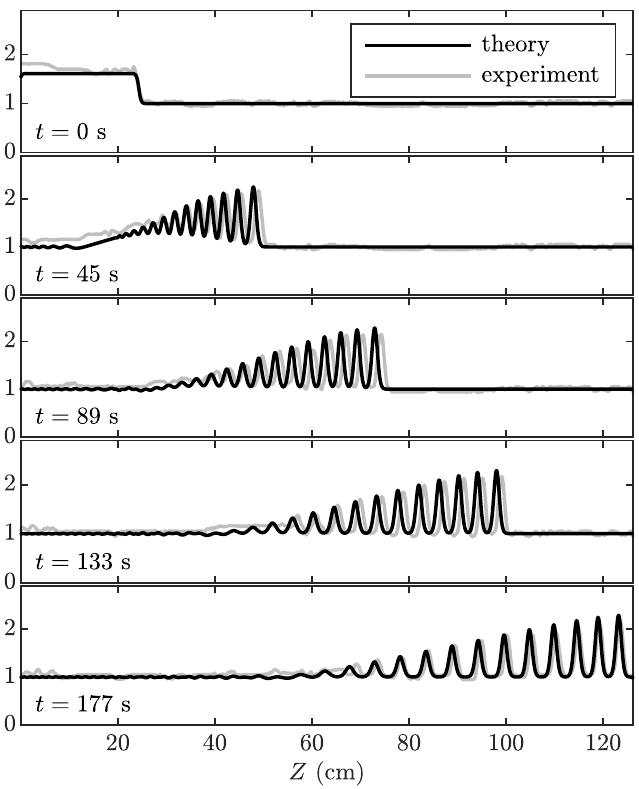}}
  \caption{Comparison of numerical (dark) and experimental (light)
    normalised conduit diameter evolution.  In both (a) and (b), the
    experimental and simulation initial box height $a_m = 1.6$ is
    determined from conduit measurements and the experimental initial
    box width of $\SI{25}{\centi\meter}$ is obtained by the boundary
    control method \citep{anderson_controlling_2019}.  Experimental
    diameter is extracted from images in figure~\ref{fig:expt_setup}(b).
    (a) Measured, nominal experimental parameters in the caption of
    figure~\ref{fig:expt_setup} are used to determine the length
    $R_0/\sqrt{8\epsilon} = \SI{1.6}{\milli\meter}$, time
    $R_0/(U_0\sqrt{8\epsilon}) = \SI{1.17}{\second}$ scales, and
    nondimensional box width $w = 156$ for the numerical simulation.
    (b) Same as (a) except the interior and exterior viscosities are
    fitted, determining the length
    $R_0/\sqrt{8\epsilon} = \SI{1.8}{\milli\meter}$, time
    $R_0/(U_0\sqrt{8\epsilon}) = \SI{1.13}{\second}$ scales, and
    nondimensional box width $w = 137$ for the numerical simulation. }
  \label{fig:expt_numerics_evolve}
\end{figure}

The experimental diameter profiles $D(Z,T)$ reported in figure
\ref{fig:expt_numerics_evolve} have been extracted from the images in
figure~\ref{fig:expt_setup}(b) per the description in Section
\ref{sec:Methods}.  Because of the large aspect ratio inherent in
these long wave dynamics, low image resolution in the transverse
direction implies that $\SI{8}{\pixel} \le D \le \SI{20}{\pixel}$.
However, from the box and wave cameras, we have much more accurate
measurements of the conduit diameter near the bottom and top of the
apparatus for both the equilibrium conduit diameter
($\SI{2}{\milli\meter}$) and the diameter of the box
($\SI{3.2}{\milli\meter}$).  We use these measurements to normalise
the pixel data via the linear transformation
$D'(Z,T) = \alpha D(Z,T) + \beta$.  The coefficients
$\alpha = \SI{0.11}{\per\pixel}$ and $\beta = 0.30$ are chosen so that
the mean equilibrium conduit is $\overline{D'} = 1$ and the box
diameter satisfies $D' = 3.2/2 = 1.6$.  This normalisation effectively
registers the low resolution data in figure~\ref{fig:expt_setup} with
the high resolution measurements from the other cameras.  The profiles
$D'(Z,T)$ are reported in figure~\ref{fig:expt_numerics_evolve} with
the lighter gray curves.

While the experiment gives rise to 14 solitary waves, the numerics
produce 16.  But the lead solitary wave diameter is only 4.5\% larger
than experiment.  The numerical evolutionary timescale is somewhat
slower than the experimental one, which is consistent with previous
measurements of large amplitude solitary waves that were found to
propagate faster than the conduit equation's solitary wave
speed-amplitude relation \eqref{eq:2} predicts
\citep{olson_solitary_1986,maiden_observation_2016}.

In figure~\ref{fig:expt_numerics_evolve}(b), we utilise the same
measured parameters as in (a) except we fit the interior
$\mu^{(i)} = \SI{6.88d-2}{\pascal\second}$ and exterior
$\mu^{(e)} = \SI{0.9}{\pascal\second}$ viscosities by increasing the
nondimensional length scale by the factor $16/14$ and reducing the
nondimensional time scale by the factor $0.97$.  This particular
increase in length scale derives from the predicted linear scaling of
the number of solitary waves by the initial box width.  Indeed,
figure~\ref{fig:expt_numerics_evolve}(b) exhibits precisely 14
solitary waves from both numerics (nondimensional initial box width
$w = 137$) and experiment.  The increased length scale and slightly
reduced time scale lead to significantly improved solitary wavetrain
evolution when compared with experiment.  Remarkably, at the final
reported time $t = \SI{177}{\second}$, the numerical and experimental
normalised diameter profiles are almost indistinguishable for the 11
largest solitary waves in the solitary wavetrain.

For the fit, the exterior viscosity is reduced by 10\%.  The high
viscosity glycerine utilised for the exterior fluid is extremely
sensitive to even small amounts of interior fluid mass diffusion from
the conduit.  A 10\% reduction from its nominal value is certainly
possible.  The interior viscosity's fitted value is approximately 39\%
larger than its measured value, which is a bit more than expected.
However, we have not accounted for uncertainty in the volumetric flow
rate or fluid densities.  Moreover, the conduit equation is a
long-wave approximation of the full two-fluid, free-boundary dynamics
that is valid in the small viscosity ratio
$\epsilon = \mu^{(i)}/\mu^{(e)}$ regime
\citep{lowman_dispersive_2013}.  For these experiments, the measured
value of this ratio is $\epsilon = 0.049$ and the fitted value is
$\epsilon = 0.076$.  Despite these reasonably small nondimensional
values, a number of higher order effects, e.g., inertia and the
finite-sized boundary \citep{lowman_dispersive_2013}, could be
influencing the dynamics on the long timescales considered---the
nondimensional final time is approximately 150 in both figures
\ref{fig:expt_numerics_evolve}(a) and
\ref{fig:expt_numerics_evolve}(b).  For these reasons, we find the
comparison between experiment and the conduit equation reported in
figure~\ref{fig:expt_numerics_evolve}(b) to be credible, strong
evidence for both the conduit equation as an accurate model of viscous
fluid conduit dynamics and the efficacy of the solitary wave
resolution method.


\section{Conclusion}\label{sec:conclusion}

We have derived explicit formulae accurately predicting the asymptotic
number and amplitude distribution of solitary waves that emerge in
long time from a localised, slowly varying initial disturbance for the
conduit equation.  Our analytical approach to the solitary wave
fission problem is based upon Whitham modulation theory.  The
predictions have been quantitatively verified with experiments on the
interfacial dynamics of two high-contrast, viscous fluids.  While the
solitary wave resolution method utilised here was previously developed
by \cite{el_asymptotic_2008} for the Serre-Green-Naghdi equations
modeling large amplitude, shallow water waves, we have identified two
new, universal predictions for box-shaped initial disturbances: i) the
asymptotic number of solitary waves is linearly proportional to 
box width, ii) the asymptotic, normalised cumulative distribution
function for the solitary wave amplitudes is independent of box width.
Universal here means that these predictions apply to a broad class of
dispersive hydrodynamic equations \eqref{eq:16}, not just the conduit
equation.

Our experiments are the first that validate the solitary wave
resolution method and we find that the physical evolution of viscous
conduit profiles is well captured by the approximation, particularly
for large width disturbances.  All observed solitary wave counts are
within 1--2 waves of their expected values and within 10\% relative
error for disturbances producing at least twelve waves.  Amplitude
distribution predictions agree quantitatively with experiment and
demonstrate the necessity of going beyond the standard weakly
nonlinear KdV model as it applies to the viscous fluid conduit system.
The conduit's observed, full spatio-temporal evolution exhibits
remarkable fidelity to numerical simulations of the conduit equation
when the two fluids' viscosities are appropriately fitted to
effectively account for a variety of uncertainties and higher order
effects in the experiment.  When this work is considered in
conjunction with the large variety of previous experiments on viscous
fluid conduits involving solitary waves
\citep{olson_solitary_1986,scott_observations_1986,whitehead_wave_1988,helfrich_solitary_1990,lowman_interactions_2014},
wavebreaking \citep{anderson_controlling_2019}, rarefaction waves, and
dispersive shock waves
\citep{maiden_observation_2016,maiden_hydrodynamic_2018}, the
analytically tractable conduit equation, and the effectiveness of
Whitham modulation theory further bolster the claim that the viscous
fluid conduit system provides both an ideal laboratory environment and
mathematical modelling framework in which to examine dispersive
hydrodynamics and nonlinear dispersive wave dynamics more generally
\citep{lowman_dispersive_2013,maiden_modulations_2016}.

Largely due to the paucity of analytical tools for studying strongly
nonlinear wave dynamics, researchers have focused primarily on
integrable models such as the KdV and modified KdV (mKdV) equations to
obtain physical predictions for the soliton fission problem.  A case
in point is the field of internal ocean waves where strongly nonlinear
solitary waves are known to be prevalent yet can only be explained
with fully nonlinear models \citep{helfrich_long_2006}.  Because the
solitary wave resolution method is agnostic to integrable structure, a
promising application direction is the fully nonlinear
Miyata-Camassa-Choi (MMC) equations for two stratified fluid layers
\citep{miyata_internal_1985,choi_fully_1999}.  The MMC equations have
all the necessary ingredients to apply the solitary wave resolution
method \citep{esler_dispersive_2011}.

These results quantify an extension of the soliton resolution
conjecture from integrable systems to non-integrable systems that,
often times, more closely model physical systems.  The conjecture
states that localised initial conditions to nonlinear, dispersive wave
equations generically evolve in long time toward a rank-ordered train
of solitary waves separated from small amplitude, dispersive waves
\citep{segur_korteweg-vries_1973,schuur_emergence_1986,deift_collisionless_1994}.
Here, we have formally derived quantitative measures of the solitary
wave component, which typically dominates the long-time outcome in
problems with large-scale, localised initial data.  Predicting
properties of the accompanying small amplitude dispersive waves is the
next step toward quantifying an extension of the full conjecture to
non-integrable systems.

These promising results for solitary wave fission in the model conduit
system provide inspiration for future studies on this problem in other
complex systems.

This work was supported by NSF DMS-1255422 and DMS-1517291 (M.\ A.\
H.), the NSF GRFP (M.\ D.\ M.), NSF EXTREEMS-QED DMS-1407340 (E.\ G.\
W.), and EPSRC grant EP/R00515X/2 (G.\ A.\ E.).

\bibliographystyle{jfm} \bibliography{bib_conduit}

\end{document}
